\documentclass[lettersize,journal]{IEEEtran}
\usepackage{amsmath,amsfonts}
\usepackage{algorithmic}
\usepackage{algorithm}
\usepackage{array}
\usepackage[caption=false,font=normalsize,labelfont=sf,textfont=sf]{subfig}
\usepackage{textcomp}
\usepackage{amsmath}
\usepackage{stfloats}
\usepackage{url}
\usepackage{booktabs}
\usepackage{verbatim}
\usepackage{graphicx}
\usepackage{cite}
\usepackage{listings}
\usepackage{amsmath,amsfonts,color}
\usepackage{graphicx}
\usepackage{amsfonts}
\usepackage{array}
\usepackage{amssymb}
\usepackage{helvet}
\usepackage{color}
\usepackage{bm}
\usepackage{float}
\usepackage{cite}
\usepackage{multirow}
\usepackage{amssymb}
\usepackage[caption=false,font=normalsize,labelfont=sf,textfont=sf]{subfig}
\usepackage{amsmath,amsfonts}
\usepackage{algorithmic}
\usepackage{algorithm}
\usepackage{textcomp}
\usepackage{stfloats}
\usepackage{url}
\usepackage{verbatim}
\usepackage{graphicx}
\usepackage{cite}
\usepackage{amssymb}
\usepackage{makecell}
\usepackage{mathrsfs}
\usepackage{url}

\newcommand{\RNum}[1]{\uppercase\expandafter{\romannumeral #1\relax}}

\newcounter{MYtempeqncnt}

\begin{document}

\title{AirGuard: UAV and Bird Recognition  Scheme for Integrated Sensing and Communications System}
\author{Hongliang Luo, Zhonghua Chu, Tengyu Zhang,   Chuanbin Zhao, Bo Lin, and Feifei Gao,~\IEEEmembership{Fellow,~IEEE}
        % <-this % stops a space
\thanks{
	The current version of this manuscript has been accepted by IEEE Journal on Selected Areas in Communications. }
\thanks{H. Luo, Z. Chu, T. Zhang, C. Zhao, B. Lin and F. Gao are with Department of Automation, Tsinghua University, Beijing 100084, China (email: luohl23@mails.tsinghua.edu.cn; c3zhua@163.com; zhang-ty22@mails.tsinghua.edu.cn; zcb23@mails.tsinghua.edu.cn;  linb20@mails.tsinghua.edu.cn; feifeigao@ieee.org).
}% <-this % stops a space
%\thanks{Manuscript received April 19, 2021; revised August 16, 2021.}
}



\maketitle

\begin{abstract}
In this paper, we propose an unmanned aerial vehicle (UAV) and bird recognition scheme with signal processing and deep learning for integrated sensing and communications (ISAC) system.
We first provide the basic scene of   low-altitude targets monitoring, and  \textcolor{black}{formulate} the  motion equations  and echo signals  for  UAVs and  birds. 
Next, we extract the centralized micro-Doppler (cmD) spectrum and the high resolution  range profile (HRRP) of the low-altitude target from the echo signals.
Then we design a dual feature fusion enabled low-altitude target recognition network with convolutional neural network (CNN), which employs both the images of cmD  spectrum  and HRRP  as inputs to jointly
distinguish between UAV and bird.
\textcolor{black}{Meanwhile,  we generate   237600 cmD and HRRP image samples  to train, validate, and evaluate the designed  low-altitude target recognition network.}
The proposed    scheme  is termed as \emph{AirGuard},
whose effectiveness has been demonstrated by simulation results.

\end{abstract}


\section{Introduction}

As a coming evolution of civil aerospace commerce, low-altitude economy (LAE) refers to a comprehensive economic model that employs various air vehicles as carrier and builds multi-scenario flight activities  to drive the integrated development of low-altitude production, \textcolor{black}{security}, operation, service  and other related fields below a vertical height of one thousand meters\cite{10723207,10608169,HUANG2024100694,2023arXiv231109047J}. 
The core elements of LAE are unmanned aerial vehicle (UAV) and manned electric vertical take-off and landing (eVTOL)\cite{XIANG2024100140},  giving rise to plentiful activities such as low-altitude logistics,  inspections,  travel, surveying\cite{8795473,9666755}.

\textcolor{black}{
The number of UAVs   is rapidly increasing, and inevitably leads to many unrestricted and unregulated flight behaviors, which poses significant challenges to social security and privacy protection.}
Hence it is necessary to utilize advanced information technology to monitor all the cooperative UAVs and non-cooperative UAVs   simultaneously\cite{drones3010013,8675384,10077453}.
Meanwhile, the emerging sixth generation mobile communications (6G) network is developing the  integrated sensing and communications (ISAC) system, which can sense and monitor various information
of the cooperative  UAVs and non-cooperative UAVs with
 the orthogonal frequency division multiplexing (OFDM) communications signals\cite{202310141,2024arXiv240519925L,9040264,9606831,itu}.

Currently, a large amount of ISAC researches  focus on target detection, parameter estimation, and trajectory tracking of the UAV target. For example, Z.~Xiao~\emph{et~al.} propose a discrete Fourier transform (DFT) based parameter estimation method to obtain the angle, distance, and velocity  of  the UAV target\cite{10634583}.
 X.~Lu~\emph{et~al.} apply multiple signal classification (MUSIC)  algorithm  to estimate the position and velocity of the UAV target~\cite{2024arXiv240412705L}.
R.~Liu~\emph{et~al.}  test the performance of    ISAC system in locating and tracking multiple UAVs and other targets\cite{10286534}.
R.~Li~\emph{et~al.}  improve the  sensing  performance   by optimizing target assignment and resource allocation among multiple base stations (BSs)\cite{10824972}.
Although these works  conduct sufficient research on  locating and tracking UAVs,
all of them overlook a practical problem that 
 the birds   would also be illuminated by  BS and thus  cause   echoes. \textcolor{black}{Then   serious false alarms will be disgustingly activated  for UAV detection and  monitor.}

To deal with these false alarms, a natural idea is to distinguish between UAVs and birds, which is termed as  \emph{UAV and Bird Recognition} problem.
Recently, J.~Wei~\emph{et~al.}  propose a UAV feature extraction method to obtain the rotor micro-Doppler (rmD) feature caused by UAV's paddle rotation, and 
 the sinusoidal envelope characteristics exhibited in the time-Doppler spectrum are expected to be  used for the  identification of rotary UAV\cite{2024arXiv240816415W}. 
D.~Ma~\emph{et~al.}   evaluate the performance of UAV identification using various time-division duplex (TDD) patterns under 
the fifth generation mobile communications (5G) network\cite{10625724}.
Although these works preliminarily analyze the UAV's micro-Doppler (mD) effect with ISAC system, 
both of them do not  distinguish between UAVs and birds.

In traditional radar areas, some  researchers    utilize  macroscopic movements and   kinematic features such as observation trajectory, velocity, acceleration, etc., of the tracked targets \textcolor{black}{to distinguish between UAVs and birds}\cite{6875676,9282965,7060321,9925778}.
Some other researchers utilize the  mD effect caused by UAV's rotors \textcolor{black}{to recognize UAVs and birds}. 
 For example, C.~Bennett~\emph{et~al.} apply fast Fourier transform (FFT) to obtain the mD spectrum of the target, and then detect the presence and the  number of symmetrical peaks in each frame of  mD spectrum 
\textcolor{black}{to differentiate UAVs from other targets}\cite{9266702}. 
H.~Dale~\emph{et~al.}  visualize the FFT based 
 mD spectrum  as an image, employ various deep learning networks to extract the hidden features from the mD image, and   utilize these hidden features  
\textcolor{black}{to classify UAVs and other targets}\cite{9455181}. 
S. Rahman~\emph{et~al.} apply short time Fourier transform (STFT) to extract the mD spectrum of the target, and then  recognize three different types of UAVs through GoogLeNet\cite{Rahman2020MultipleDC}.
Besides, B.~Oh~\emph{et~al.} perform empirical mode decomposition (EMD)  on radar echoes to obtain a set of oscillation signals, extract statistical features from them, and  utilize support vector machine (SVM) \textcolor{black}{to identify UAVs and birds}\cite{8239598}.
Moreover, as another important feature, high resolution range profile (HRRP) can reflect the distribution of scattering intensity of  the extended target along the radar's line of sight, and  usually contains the key structural features of the target\cite{4490100,5746649,1634818}.
Hence HRRP can also be used to distinguish between UAVs and birds.
For example,  W.~Jiang~\emph{et~al.} utilize long short term memory (LSTM) network to process the  HRRP feature and convolutional neural  network (CNN) to process the mD feature, thus realizing  UAVs and other targets recognition\cite{rs16152710}.
Although these works \cite{9266702,9455181,Rahman2020MultipleDC,8239598,rs16152710} provide plentiful technical reserves for UAV and bird recognition, all of them  cannot be directly applied to ISAC system.

In this paper, we  propose a dual feature fusion enabled UAV and bird recognition scheme with signal processing and deep learning for ISAC system.
The contributions of this paper are summarized as follows.

\begin{itemize}
	
\item  We  \textcolor{black}{formulate} the motion equations and \textcolor{black}{characterize} the  echo channels of UAVs and birds by processing their 3D  mesh files and analyzing their physical motion laws.
\textcolor{black}{Then we compute the expressions for the echo signals of UAVs and birds under   ISAC system.}

\item We develop a low-altitude target feature extraction method for ISAC system, in which we utilize grouped DFT to obtain the centralized micro-Doppler (cmD) spectrum of the target, and utilize DFT estimation and  filtering algorithm to obtain the HRRP of the target.

\item We design a \textcolor{black}{dual feature fusion enabled} low-altitude target recognition network with CNN, which employs both the images of cmD  spectrum  and HRRP  as inputs to jointly
recognize the UAVs and birds.

\item \textcolor{black}{We generate a UAV and bird echo signal dataset, and then  generate 237600 cmD and HRRP image samples   to train, validate, and evaluate the designed  low-altitude target recognition network.}

\item The generated dataset and the  proposed scheme  are termed as \emph{AirGuard},  
whose effectiveness has been demonstrated by simulation results.

\end{itemize}

The remainder of this paper is organized as follows.
In Section \RNum{2}, we  construct the basic scene of ISAC system monitoring low-altitude targets.
In Section \RNum{3}, we \textcolor{black}{formulate} the echo signals of  UAVs and birds   under   ISAC system.
In Section~\RNum{4}, we propose a low-altitude target recognition scheme with signal processing and deep learning.
Simulation results and conclusions are given in Section~\RNum{5} and Section~\RNum{6}.

\begin{figure}[!t]
	\centering
	\includegraphics[width=76mm]{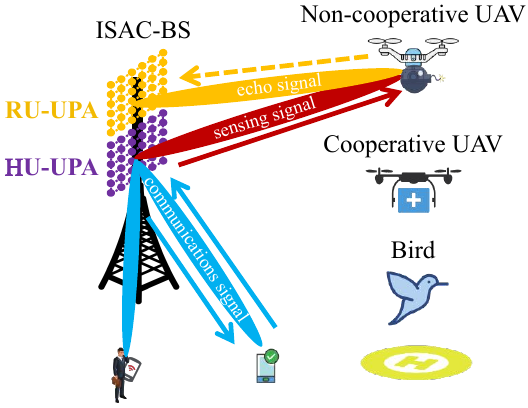}
	\caption{ISAC scenario for low-altitude target monitoring.}
	\label{fig_1}
\end{figure}

\emph{Notation}:
Lower-case and upper-case boldface letters $\mathbf{a}$ and $\mathbf{A}$ denote a vector and a matrix; $\mathbf{a}^*$, 
$\mathbf{a}^T$ and $\mathbf{a}^H$ denote the conjugate,  transpose and  conjugate transpose of  $\mathbf{a}$, respectively;
$\mathbf{a}[n]$  denotes the $n$-th element of the vector $\mathbf{a}$;
$\mathbf{A}[{i,j}]$ denotes the $(i,j)$-th element of the matrix $\mathbf{A}$; $\mathbf{A}[i_1:i_2,:]$ is the submatrix composed of all columns elements in rows $i_1$ to $i_2$ of matrix $\mathbf{A}$;
$\mathbf{A}[:,j_1:j_2]$ is the submatrix composed of all rows elements in columns $j_1$ to $j_2$ of matrix $\mathbf{A}$;
$\otimes$ denotes the Kronecker product;
$\mathbb{S}$ denotes a set; 
${\rm card}(\mathbb{S})$ represents the number of elements in set $\mathbb{S}$;
$\varnothing$ denotes an empty set; 
$\cap$ and $\cup$ respectively represent the intersection and union of sets;
$\mathbb{N}^+$ denotes the positive integer space;
$||\cdot||_2$  represents the Euclidean norm.

\section{System Model}

In this section, 
we provide the basic scenario of 
monitoring low-altitude targets with ISAC system,
and derive the echo channel model for the extended targets.

\subsection{ISAC Scene and BS Model}

Fig.~1 depicts the basic scenario of   monitoring low-altitude targets with ISAC system. Generally, the  cooperative UAVs are obliged to  report their positions, velocities,  identification numbers and other information in real-time. 
Meanwhile, the BS can transmit sensing detection signals, receive and process the sensing  echo signals to obtain various information of cooperative UAVs, non-cooperative UAVs,  and  birds.
\textcolor{black}{For conciseness,  cooperative UAVs and non-cooperative UAVs are uniformly named as UAVs, and then the  BS should  recognize UAVs and birds by   processing the sensing echo signals.}

The ISAC  BS employs massive MIMO arrays and OFDM signals operating in the mmWave frequency bands to realize both  wireless communications and low-altitude target monitoring.
As shown in Fig.~1, the BS is equipped with one {hybrid unit uniform planar array (HU-UPA)} and one {radar  unit uniform planar array (RU-UPA)}.
The HU-UPA is responsible for transmitting downlink communications signals and receiving uplink communications signals, as well as transmitting downlink sensing  signals. The RU-UPA is   responsible for receiving sensing echo signals. 
Meanwhile,
HU-UPA and RU-UPA are each equipped with $N_H=N^{x}_{H}\times N_{H}^z$ and $N_R=N_{R}^x\times N_{R}^z$ antenna elements, respectively,  and are both vertically mounted on the 2D plane $y = 0$. 
\textcolor{black}{The antenna spacing between the antennas  along  x-axis and z-axis are 
$d_x = d_z =  d = \frac{\lambda}{2}$,
where $\lambda = c/f_0$ is the wavelength, $f_0$ is   the operating frequency of the system, and $c$ represents the speed of light.} 
Without loss of generality, 
we   assume that HU-UPA and  RU-UPA
are co-located at the origin of the coordinate system  and are parallel to each other,  such that they can see the  low-altitude  targets at the same propagation directions\footnote{Since  the  communications and sensing distance is much longer than the protection distance between  HU-UPA and RU-UPA, \textcolor{black}{we can consider} that HU-UPA and RU-UPA are located in the same position.}\cite{9898900}.
%By balancing the hardware costs and system performance,
We also assume that  HU-UPA and RU-UPA employ the phase shifter (PS)  based hybrid
 hardware architecture, in which a total of $N_{H,RF}\ll N_H$ and $N_{R,RF}\ll N_R$ radio frequency (RF) chains are deployed, respectively, and each antenna is connected to one PS.

Suppose that  ISAC system emits   OFDM signals with $M$ subcarriers,
where the  lowest  frequency and the subcarrier  interval  are $f_0$ and $\Delta f$, respectively. 
Then the transmission bandwidth is $W=M\Delta f$, and 
the frequency of the $m$-th subcarrier  is $f_m=f_0+m\Delta f$,  $m=0,1,...,M-1$.  
Define one OFDM frame specifically used for low-altitude target recognition as
 one \emph{target recognition frame (TCF)}, which will enable idle or dedicated time-frequency resources of the BS across multiple OFDM frames in 5G new radio (NR) standard\cite{2024arXiv240816415W}.
For clarity, we directly consider that one \textcolor{black}{TCF} contains $N$ consecutive OFDM symbols, where  
the time interval between the adjacent OFDM symbols is  $T_s = T'_s+T_g$, 
with $T'_s = \frac{1}{\Delta f}$ and $T_g$
being the OFDM symbol duration and guard interval, respectively\cite{4570206}.

Let us employ the spherical coordinate $(r,\theta,\phi)$  in 3D space,
where $r$  represents the polar distance with   range  $r \geq 0$,
$\theta$ represents  the horizontal angle with  range  $0^\circ \leq \theta \leq 180^\circ$, and $\phi$  represents the pitch angle with   range  $-90^\circ \leq \phi \leq 90^\circ$. 
\textcolor{black}{Noting that  BS is located at the origin of the coordinate system, we  denote the service area of BS as 
$\{(r,\theta,\phi)|r_{min}\leq r \leq r_{max},\theta_{min}\leq \theta \leq \theta_{max},\phi_{min}\leq \phi \leq \phi_{max}\}$.}
Then we assume that all communications users and low-altitude targets are within this service area. 
Since the co-existence design of communications and sensing has been extensively discussed in many existing works, and the purpose of this work is  low-altitude target recognition, we will ignore the communications stage and focus on how to use ISAC system to realize   UAV and bird recognition.

\subsection{Echo Channel Model of the Extended Target}

The ISAC BS first transmits sensing detection signals through HU-UPA, which are then scattered by the low-altitude targets  and cause echoes. These  echo signals will be received by  RU-UPA at the BS.  
To observe  the detailed features of low-altitude targets, UAVs and  birds should be modeled as \emph{extended targets} that can  be represented by the \emph{multiple scattering points model}.
In  multiple scattering points model, 
 the extended target could be  modeled as a combination of several spatially adjacent \emph{point targets}, and  the echo channel of the extended target could be modeled as the  sum of the echo channels of all corresponding point targets\cite{9947033}.

Let us first consider a point target with  motion parameters  $\{r, \theta, \phi, v_{r}, \omega_{\theta}, \omega_{\phi}\}$,
 which 
 represent the distance, horizontal angle, pitch angle, radial velocity, horizontal angular velocity, and pitch angular velocity relative to the BS, respectively.
Then the  echo channel matrix of this point target on the $m$-th subcarrier of the $n$-th OFDM symbol can be represented as 
\begin{equation}
	\begin{split}
		\begin{aligned}
			\label{deqn_ex1a}
\mathbf{H}_{n,m}^{\rm point} \!=\! \alpha  e^{\!-\!j\! \frac{4\pi\! f_m  r}{c}}\!
			e^{j4\pi f_m\!\frac{v_{r}nT_s}{c}} \! 
			\mathbf{a}_{\!R}(\Psi_{n},\Omega_{n})
			\mathbf{a}^T_{\!H}(\Psi_{n},\Omega_{n}),
		\end{aligned}
	\end{split}
\end{equation}
where $\alpha$ represents the channel  fading factor, 
$\Psi_{n} = \Psi(\theta_{n},\phi_{n})=\cos \phi_{n}\!\cos \theta_{n} = \cos (\phi - \omega_{\phi} nT_s) \cos (\theta - \omega_{\theta} nT_s)$ represents the horizontal spatial-domain direction, 
$\Omega_{n}=\Omega (\theta_{n},\phi_{n})=\sin \phi_{n}=\sin (\phi - \omega_{\phi} nT_s)$ represents the pitch spatial-domain direction, 
$\mathbf{a}_{R}(\Psi,\Omega)$ and $\mathbf{a}_{H}(\Psi,\Omega)$ are the array  steering vectors for the spatial-domain direction  $(\Psi,\Omega)$ of RU-UPA and HU-UPA with the form\cite{7523373,9573459}
\begin{align}
\mathbf{a}_{R}(\Psi,\Omega) &=
			\mathbf{a}_{R}^x(\Psi)\otimes \mathbf{a}_{R}^z(\Omega)
			\in \mathbb{C}^{N_R\times 1},\\
\mathbf{a}_{H}(\Psi,\Omega) &=
\mathbf{a}_{H}^x(\Psi )\otimes \mathbf{a}_{H}^z(\Omega)
	 \in \mathbb{C}^{N_H\times 1}.
\end{align}
Here, $\otimes$ denotes the Kronecker product, and 
\begin{align}
	\!\!\mathbf{a}_{R}^x(\Psi)&\!=\![1,e^{j\frac{2\pi f_0d\Psi}{c}},...,e^{j\frac{2\pi f_0d\Psi}{c}(N_{R}^x-1)}]^T \!\!\in\! \mathbb{C}^{N_{R}^x\times 1},\\
	\!\!\mathbf{a}_{R}^z(\Omega)&\!=\! [1,e^{j\frac{2\pi f_0d\Omega}{c}},...,e^{j\frac{2\pi f_0d\Omega}{c}(N_{R}^z-1)}]^T \!\! \in\! \mathbb{C}^{N_{R}^z\times 1},\\
\!\!\mathbf{a}_{H}^x(\Psi)&\!=\![1,e^{j\frac{2\pi f_0d\Psi}{c}},...,e^{j\frac{2\pi f_0d\Psi}{c}(N_{H}^x-1)}]^T \!\!\in\! \mathbb{C}^{N_{H}^x\times 1},\\
\!\!\mathbf{a}_{H}^z(\Omega)&\!=\! [1,e^{j\frac{2\pi f_0d\Omega}{c}},...,e^{j\frac{2\pi f_0d\Omega}{c}(N_{H}^z-1)}]^T \!\! \in\! \mathbb{C}^{N_{H}^z\times 1}.
\end{align}

\begin{figure*}[!t]
	\centering
	\includegraphics[width=180mm]{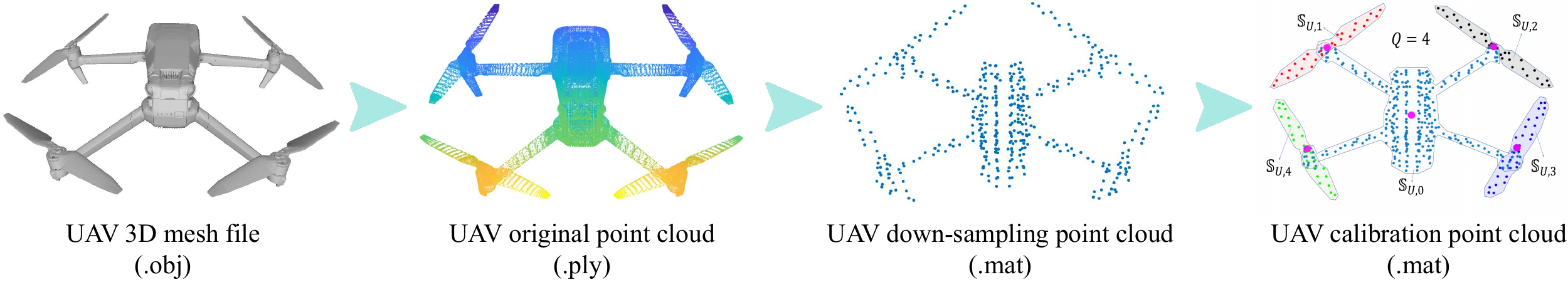}
	\caption{Process diagram to obtain the UAV calibration point cloud.}
	\label{fig_1}
\end{figure*}

Note that  Eq.~(1) can be equivalently rewritten as the following set of equations
\begin{equation}
\left\{
\begin{aligned}
&r_n = r - v_r n T_s, \\
&\theta_n = \theta - \omega_{\theta} nT_s,\\
&\phi_n = \phi - \omega_{\phi} nT_s, \\
&\Psi_n = \cos \phi_{n}\!\cos \theta_{n},\\
&\Omega_n  = \sin \phi_{n},\\
&\mathbf{H}_{n,m}^{\rm point} =\! \alpha  e^{\!-\!j\! \frac{4\pi\! f_m  r_n}{c}}\!\mathbf{a}_{\!R}(\Psi_{n},\Omega_{n})
			\mathbf{a}^T_{\!H}(\Psi_{n},\Omega_{n}).
\end{aligned}
\right.
\end{equation}
Since equations in (8) contain the motion   parameters of the point target,  target detection and parameter estimation tasks are typically implemented based on equations in (8) or the simplified form\cite{10477890,10854508}.
More importantly, 
  target detection and parameter estimation tasks  typically require only a small amount of symbol observation time,  and hence 
 the velocities of the point target, i.e., $v_r$, $\omega_{\theta}$, and $\omega_{\phi}$ are considered constant in  existing ISAC researches. Then the  motion equations of the point target in  (8), i.e. $r_n = r - v_r n T_s$, $\theta_n = \theta - \omega_{\theta} nT_s$, and $\phi_n = \phi - \omega_{\phi} nT_s$ are  linear equations.

However, when the ISAC system attempts to distinguish between UAVs and birds, it usually requires a longer symbol observation time.
Then some scattering points in UAVs and birds will exhibit nonlinear motion such as rotation, jitter, etc. Therefore, let us rewrite the equations in (8) as
\begin{equation}
\left\{
\begin{aligned}
&(r_n,\theta_n,\phi_n) = g(r,\theta,\phi,v_r,\omega_{\theta},\omega_{\phi},n),\\
&(\Psi_n,\Omega_n)= (\cos \phi_{n}\!\cos \theta_{n},\sin \phi_{n}),\\
&\mathbf{H}_{n,m}^{\rm point} =\! \alpha  e^{\!-\!j\! \frac{4\pi\! f_m  r_n}{c}}\!\mathbf{a}_{\!R}(\Psi_{n},\Omega_{n})
			\mathbf{a}^T_{\!H}(\Psi_{n},\Omega_{n}),
\end{aligned}
\right.
\end{equation}
where $g(\cdot)$ represents the physical motion function.
By splitting $(r_n,\theta_n,\phi_n) = g(r,\theta,\phi,v_r,\omega_{\theta},\omega_{\phi},n)$ into several mixed linear and nonlinear motion update steps, we can use the equations in (9) to more accurately characterize the motion details of low-altitude targets.

Based on the equations in (9), 
for an extended target consisting of $L$ scattering points, the sensing echo channel  of this extended target on the $m$-th subcarrier  of the $n$-th OFDM symbol can be expressed as
\begin{equation}
	\begin{split}
		\begin{aligned}
			\label{deqn_ex1a}
\!\!\!\!\! \mathbf{H}_{n,m} &=\sum_{l=1}^{L} \mathbf{H}_{l,n,m}^{\rm point}\\&=
\sum_{l=1}^{L}\alpha_l  e^{\!-\!j\! \frac{4\pi\! f_m  r_{l,n}}{c}}\!\mathbf{a}_{\!R}(\Psi_{l,n},\Omega_{l,n})
			\mathbf{a}^T_{\!H}(\Psi_{l,n},\Omega_{l,n}),
		\end{aligned}
	\end{split}
\end{equation}
where $r_{l,n}, \Psi_{l,n},\Omega_{l,n}$ indicate the position information of the $l$-th scattering point of this extended target at the $n$-th  symbol.

\section{\textcolor{black}{Echo Signals   of UAVs and Birds}}

In this section, we  \textcolor{black}{formulate} the motion equations and \textcolor{black}{characterize} the  echo channels of UAVs and birds by processing their 3D  mesh files and analyzing their physical motion laws.
\textcolor{black}{Then we compute the expressions for the echo signals of UAVs and birds under   ISAC system.}

\subsection{Motion Equations and Echo Channels of the UAVs}

As shown in Fig.~2,  \textcolor{black}{we  import the 3D  mesh file of the UAV (.obj file\footnote{\textcolor{black}{The ``.obj file'' is a plain-text file format used to define 3D geometries, including vertices, textures, and normals, commonly utilized in 3D modeling and graphics applications. The ``.ply file'' is a file format used to store 3D geometry data, including points, faces, and properties such as color or normals, often used in 3D scanning and modeling.  The ``.mat file'' is a binary file format used by MATLAB to store variables, arrays, and data structures for efficient data management and analysis.}}) into the MeshLab software,}
an open source system that is  portable and extensible for processing and editing the unstructured large 3D triangular meshes\cite{mesh1}. 
To obtain the effective  point cloud that  represents the skeleton of the UAV, we use  the Butterfly Subdivision Surface algorithm   to interpolate and refine the 3D  mesh of the UAV\cite{dyn1990butterfly}. 
Then we export the interpolated 3D  mesh as the \emph{original point cloud} (.ply file) of the UAV. 
Next, we load the original point cloud   and perform uniform down-sampling on it.
Then we manually calibrate the UAV's body and each paddle from the \emph{down-sampling point cloud} to obtain the  \textcolor{black}{\emph{calibration point cloud} (CPC)} of the UAV.  
In CPC, denote the set of scattering points
of the entire UAV as $\mathbb{S}_U$ with ${\rm card} \  (\mathbb{S}_U) = L_U$,  denote the set of scattering points of the UAV's body
as $\mathbb{S}_{U,0}$, and denote the set of scattering points of the UAV's each paddle as $\mathbb{S}_{U,1}$, ..., $\mathbb{S}_{U,Q}$, where $Q$ is the number of UAV's paddles with 
${\rm card} \  (\mathbb{S}_{U,i}) = L_{U,i}$, $i = 0,1,...,Q$.
There are $\bigcup_{i=0}^{Q} \mathbb{S}_{U,i} = \mathbb{S}_U$,  $\mathbb{S}_{U,i} \cap \mathbb{S}_{U,j} = \varnothing$ for $i \neq j \in \{0,1,...,Q\}$, and $\sum_{i=0}^Q L_{U,i} = L_{U}$.
\textcolor{black}{
Meanwhile, denote the position of the $l_{U,i}$-th scattering point in $\mathbb{S}_{U,i}$ as 
$\widetilde{\mathbf{p}}_{l_{U,i},n=-1}$, with $i = 0,1,...,Q$  and  $l_{U,i} = 1,...,L_{U,i}$.
Then the CPC of  the UAV can be denoted as  $\{\widetilde{\mathbf{p}}_{l_{U,i},n=-1} |  i = 0,1,...,Q; l_{U,i} = 1,...,L_{U,i}\}$,
whose center position is roughly located at the origin of the coordinate system.}

\begin{figure}[!t]
	\centering
	\includegraphics[width=90mm]{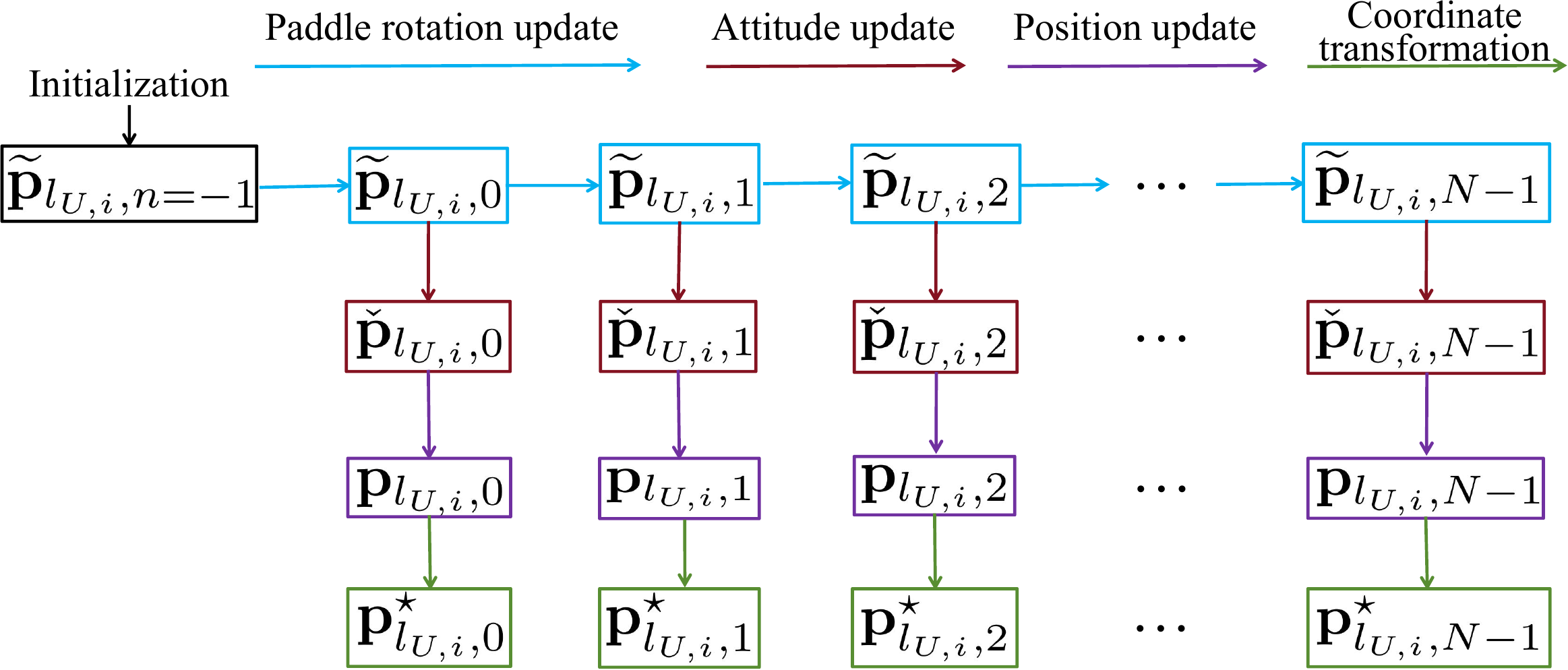}
	\caption{Update process of UAV motion equations.}
	\label{fig_1}
\end{figure}

\textcolor{black}{Next, we need to analyze the motion laws of the UAV to formulate its motion equations.} 
One of the main characteristics of UAV motion is the process of paddle rotation, which can be formulated as the scattering points in each paddle rotating around a corresponding rotation axis. 
Let us consider the $0$-th OFDM symbol  as the initial moment. 
We  iteratively update  the paddle rotation from UAV's CPC to obtain the \emph{paddle rotation update point cloud} (PRUPC) as $\{\widetilde{\mathbf{p}}_{l_{U,i},n} |  i = 0,1,...,Q; l_{U,i} = 1,...,L_{U,i}\}$,  $n=0,...,N-1$. 
Specifically, denote the initial phase  and rotation frequency  of the $i$-th paddle  as  $\beta_{U,i,0}$ and $F_{U,i}$, respectively,
and denote the phase of paddle   at the $n$-th symbol  as $\beta_{U,i,n}$,  $i=1,...,Q$.
We manually designate two points for
the $i$-th paddle as the rotation axis, which are denoted as $\mathbf{p}^a_{U,i}$ and $\mathbf{p}^b_{U,i}$.   Then the unit directional vector of the 
 rotation axis   is $\mathbf{e}_{U,i} = \frac{\mathbf{p}^a_{U,i} - \mathbf{p}^b_{U,i}}{\left|\mathbf{p}^a_{U,i} - \mathbf{p}^b_{U,i}\right|_2}$. 
When $n \!=\! 0,...,N-1$, 
we calculate the phase change of the $i$-th paddle as $\Delta \beta_{U,i,n} \!=\! \beta_{U,i,0}$ for $n=0$, and $\Delta \beta_{U,i,n} = 2\pi F_{U,i} T_s$ for $n=1,...,N-1$.  
\textcolor{black}{Then we should rotate the point $\widetilde{\mathbf{p}}_{l_{U,i},n-1}$ around the $i$-th rotation axis by angle $\Delta \beta_{U,i,n}$ to obtain the updated point $\widetilde{\mathbf{p}}_{l_{U,i},n}$.
According to the Rodrigues' rotation formula\cite{dai2015euler}, 
$\widetilde{\mathbf{p}}_{l_{U,i},n}$ can be 
 represented as}
\begin{equation}
	\begin{split}
		\begin{aligned}
			\label{deqn_ex1a}
\widetilde{\mathbf{p}}_{l_{U,i},n} & \!= \mathbf{p}^a_{U,i} + \mathbf{u}_{l_{U,i},n}\cos \Delta \beta_{U,i,n} \\& \quad \! + (\mathbf{e}_{U,i} \times \mathbf{u}_{l_{U,i},n})\sin \Delta \beta_{U,i,n} \\& \quad \!+ \mathbf{e}_{U,i} (\mathbf{e}_{U,i}^T   \mathbf{u}_{l_{U,i},n})(1\!-\! \cos \Delta \beta_{U,i,n}), \!\!\! \quad i=1,...,Q, 
		\end{aligned}
	\end{split}
\end{equation}
with $\mathbf{u}_{l_{U,i},n} \!=\! \widetilde{\mathbf{p}}_{l_{U,i},n-1} \!-\! \mathbf{p}^a_{U,i}$.
Meanwhile, the UAV's body   will not change
in PRUPC, and hence there is $\widetilde{\mathbf{p}}_{l_{U,0},n} = \widetilde{\mathbf{p}}_{l_{U,0},n-1}$.

Another important characteristic of UAV motion is the  spatial attitude, typically described by three attitude angles, i.e., roll angle, pitch angle, and yaw angle. Meanwhile, due to the factors such as atmospheric turbulence and body vibration, the motion of UAV will exhibit significant jitter effect, mainly manifested as small-scale fluctuations in attitude angles\cite{9573459}.
With these motion laws, we update  the attitude change from PRUPC to obtain the  UAV's  \emph{attitude update point cloud} (AUPC)  as $\{\check{\mathbf{p}}_{l_{U,i},n} |  i = 0,1,...,Q; l_{U,i} = 1,...,L_{U,i}\}$, $n=0,...,N-1$. 
Specifically,   we denote the initial  attitude and attitude jitter matrix of the UAV as $[\gamma_{U,Y,0},\gamma_{U,P,0},\gamma_{U,R,0}]^T$ and $\mathbf{J}_{U} \in \mathbb{R}^{3\times 2}$, respectively,
and then denote the  attitude at the $n$-th symbol  as 
$[\gamma_{U,Y,n},\gamma_{U,P,n},\gamma_{U,R,n}]^T$.
When $n = 0,...,N-1$,  calculate the attitude at the $n$-th symbol time as 
$[\gamma_{U,Y,n},\gamma_{U,P,n},\gamma_{U,R,n}]^T = [\gamma_{U,Y,0},\gamma_{U,P,0},\gamma_{U,R,0}]^T + \sum_{j=0}^{n} [\Delta \! \gamma_{U,Y,n}, \Delta \! \gamma_{U,P,n},\Delta \! \gamma_{U,R,n}]^T$, where $\Delta \gamma_{U,Y,n}, \Delta \gamma_{U,P,n},\Delta \gamma_{U,R,n}$ are the Gaussian random numbers with mean $\mathbf{J}_{U}[1,1], \mathbf{J}_{U}[2,1], \mathbf{J}_{U}[3,1]$ and variance $\mathbf{J}_{U}[1,2], \mathbf{J}_{U}[2,2], \mathbf{J}_{U}[3,2]$, respectively.
Then the updated point  $\check{\mathbf{p}}_{l_{U,i},n}$ in UAV's  AUPC corresponding to the point $\widetilde{\mathbf{p}}_{l_{U,i},n}$ in PRUPC can be 
represented as 
\begin{equation}
	\begin{split}
		\begin{aligned}
			\label{deqn_ex1a}
\check{\mathbf{p}}_{l_{U,i},n} = \mathbf{R}_{Y}(\gamma_{U,Y,n}) \mathbf{R}_{P}(\gamma_{U,P,n}) \mathbf{R}_{R}(\gamma_{U,R,n}) \widetilde{\mathbf{p}}_{l_{U,i},n},
		\end{aligned}
	\end{split}
\end{equation}
\textcolor{black}{where}
\begin{equation}\color{black}
	\begin{split}
		\begin{aligned}
			\label{deqn_ex1a}
\mathbf{R}_Y(\gamma_{U,Y,n}) = 
\begin{pmatrix}
\cos  \gamma_{U,Y,n} & -\sin \gamma_{U,Y,n} & 0 \\
\sin \gamma_{U,Y,n}  & \cos  \gamma_{U,Y,n} & 0 \\
0 & 0  & 1
\end{pmatrix},
		\end{aligned}
	\end{split}
\end{equation}
\begin{equation}\color{black}
	\begin{split}
		\begin{aligned}
			\label{deqn_ex1a}
\mathbf{R}_P(\gamma_{U,P,n}) = 
\begin{pmatrix}
\cos  \gamma_{U,P,n} & 0 &  \sin \gamma_{U,P,n}   \\
0 & 1  & 0\\
- \sin \gamma_{U,P,n}  & 0 & \cos  \gamma_{U,P,n} 
\end{pmatrix},
		\end{aligned}
	\end{split}
\end{equation}
\begin{equation}\color{black}
	\begin{split}
		\begin{aligned}
			\label{deqn_ex1a}
\mathbf{R}_R(\gamma_{U,R,n}) = 
\begin{pmatrix}
1 & 0 & 0 \\
0 & \cos  \gamma_{U,R,n} & -\sin \gamma_{U,R,n}  \\
0 & \sin \gamma_{U,R,n}  & \cos  \gamma_{U,R,n}  
\end{pmatrix},
		\end{aligned}
	\end{split}
\end{equation}
\textcolor{black}{are the  attitude rotation matrices\cite{9573459}.}

\begin{figure*}[!t]
	\centering
	\includegraphics[width=180mm]{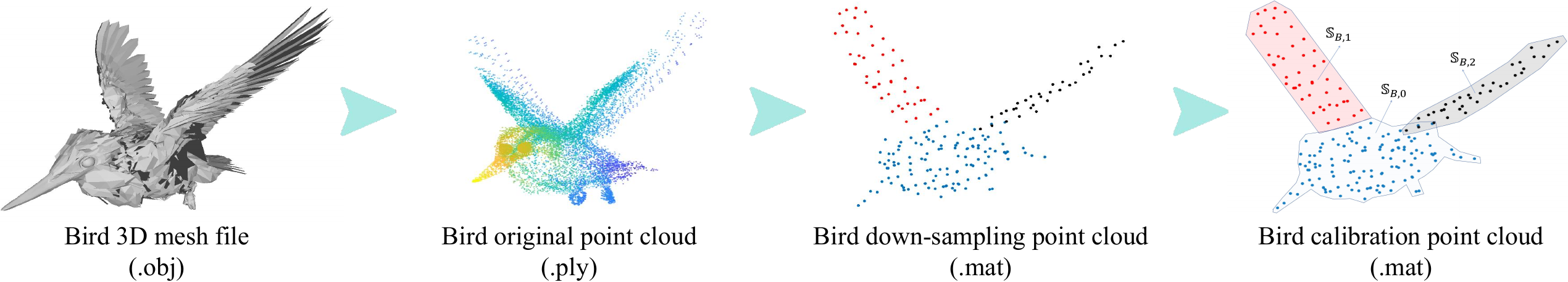}
	\caption{Process diagram to obtain the bird calibration point cloud.}
	\label{fig_1}
\end{figure*}

Next, we need to update the position change from UAV's AUPC to obtain the  UAV's 
\emph{Cartesian coordinate position point cloud} (CCPPC) as $\{{\mathbf{p}}_{l_{U,i},n} |  i = 0,1,...,Q; l_{U,i} = 1,...,L_{U,i}\}$, $n=0,...,N-1$.
Denote the initial position and motion velocity of the UAV as 
$\mathbf{p}_{U,0} = [x_{U,0},y_{U,0},z_{U,0}]^T$ and $\mathbf{v}_{U} = [v_{U,x},v_{U,y},v_{U,z}]^T$,  respectively, and  denote the position of the UAV at the $n$-th symbol  as $\mathbf{p}_{U,n} = [x_{U,n},y_{U,n},z_{U,n}]^T$.
 When $n = 0,...,N-1$,  calculate the position of the UAV at the $n$-th symbol  as $\mathbf{p}_{U,n} =\mathbf{p}_{U,0} +  nT_s  \mathbf{v}_{U}$. Then 
 the  point $\mathbf{p}_{l_{U,i},n}$
 can be represented as 
\begin{equation}
	\begin{split}
		\begin{aligned}
			\label{deqn_ex1a}
\mathbf{p}_{l_{U,i},n} = \check{\mathbf{p}}_{l_{U,i},n} + \mathbf{p}_{U,n}.
		\end{aligned}
	\end{split}
\end{equation}
Note that
the spherical coordinate $(r,\theta,\phi)$  and Cartesian coordinate $(x,y,z)$ can be converted into each other through $x=r\cos \phi \cos \theta,\quad y=r\cos \phi \sin \theta,\quad z=r\sin \phi $.
Here we   convert the  CCPPC    into the 
\emph{polar coordinate position point cloud} (PCPPC),   denoted as $\{ \mathbf{p}^{\star}_{l_{U,i},n} = ({r}_{l_{U,i},n}, {\theta}_{l_{U,i},n}, {\phi}_{l_{U,i},n}) |  i = 0,1,...,Q; l_{U,i} = 1,...,L_{U,i}\}$, $n = 0,...,N-1$.
This PCPPC 
 is  considered as the 
multiple scattering points model of the UAV at the $n$-th OFDM symbol.

Fig.~3 summarizes the update process of the UAV motion equations. 
After obtaining the multiple scattering points model $\{ \mathbf{p}^{\star}_{l_{U,i},n} = ({r}_{l_{U,i},n}, {\theta}_{l_{U,i},n}, {\phi}_{l_{U,i},n}) |  i = 0,1,...,Q; l_{U,i} = 1,...,L_{U,i}\}$ with all $N$ OFDM symbols, 
based on Eq.~(10),
 we can characterize the sensing echo channel matrix of the UAV on the  $m$-th subcarrier of the $n$-th OFDM symbol   as 
\begin{equation}
	\begin{split}
		\begin{aligned}
			\label{deqn_ex1a}
\mathbf{H}^{\rm UAV}_{n,m} = \sum_{i=0}^{Q}
 &  \sum_{l_{U,i}=1}^{L_{U,i}}
 \alpha_{l_{U,i}}  e^{\!-\!j\! \frac{4\pi\! f_m  {r}_{l_{U,i},n}}{c}}\!
\times  \\
& \mathbf{a}_{\!R}(\Psi_{l_{U,i},n},\Omega_{l_{U,i},n})
			\mathbf{a}^T_{\!H}(\Psi_{l_{U,i},n},\Omega_{l_{U,i},n}),
		\end{aligned}
	\end{split}
\end{equation}
where $\Psi_{l_{U,i},n} \!=\! \cos {\phi}_{l_{U,i},n} \cos {\theta}_{l_{U,i},n}$ and $\Omega_{l_{U,i},n} \!=\! \sin {\phi}_{l_{U,i},n}$.

\subsection{Motion Equations and Echo Channels of the Birds}

\begin{figure*}[!t]
	\normalsize
	\setcounter{MYtempeqncnt}{\value{equation}}
	\setcounter{equation}{40}
	\begin{equation}\color{black}
		\begin{split}
			\begin{aligned}
				\label{deqn_ex1a}
\Delta \beta_{B,1,n} = \left \{
\begin{array}{cl}
-2\beta_c F_B T_s &  \Delta \beta_{B,1,n-1} > 0,  \beta_{B,1,n-1} >\beta_{Bmax} \\
2\beta_c F_B T_s & \Delta \beta_{B,1,n-1} > 0, \beta_{Bmin} <  \beta_{B,1,n-1} <\beta_{Bmax}\\
2\beta_c F_B T_s & \Delta \beta_{B,1,n-1} > 0,  \beta_{B,1,n-1} <\beta_{Bmin} \\
- 2\beta_c F_B T_s & \Delta \beta_{B,1,n-1} < 0,  \beta_{B,1,n-1} >\beta_{Bmax}\\
- 2\beta_c F_B T_s & \Delta \beta_{B,1,n-1} < 0,  \beta_{Bmin} < \beta_{B,1,n-1} <\beta_{Bmax}\\
2\beta_c F_B T_s & \Delta \beta_{B,1,n-1} < 0,  \beta_{B,1,n-1}  <\beta_{Bmin}
\end{array} \right .
			\end{aligned}
		\end{split}\tag{18}
	\end{equation}
	\hrulefill % 添加一条水平线
	\vspace*{4pt}
	\setcounter{equation}{\value{MYtempeqncnt}}
	% The IEEE uses as a separator
\end{figure*}

As shown in Fig.~4, we use MeshLab to convert the 3D  mesh file of the bird into the original point cloud, and then  perform uniform down-sampling  to obtain the down-sampling point cloud.  Next we manually calibrate the bird's body and two wings to obtain the 
  CPC of the bird,  denoted as $\mathbb{S}_B$ with ${\rm card} \ (\mathbb{S}_B) = L_B$. Since bird only has two wings, we denote the set of scattering points of the bird's body
as $\mathbb{S}_{B,0}$, and denote the set of scattering points of the bird's each wing as $\mathbb{S}_{B,1}$ and $\mathbb{S}_{B,2}$, where ${\rm card} \ (\mathbb{S}_{B,i}) =L_{B,i}$, $i=0,1,2$.
There are $\bigcup_{i=0}^{2} \mathbb{S}_{B,i} = \mathbb{S}_B$,  $\mathbb{S}_{B,i} \cap \mathbb{S}_{B,j} = \varnothing$ for $i \neq j \in \{0,1,2\}$, and $\sum_{i=0}^2 L_{B,i} \!=\! L_{B}$.
\textcolor{black}{Meanwhile,
denote the position of the $l_{B,i}$-th scattering point in $\mathbb{S}_{B,i}$ as 
$\widetilde{\mathbf{p}}_{l_{B,i},n=-1}$ with $i = 0,1,2$  and  $l_{B,i} \!=\! 1,...,L_{B,i}$.
Then the CPC of  the bird can be denoted as  $\{\widetilde{\mathbf{p}}_{l_{B,i},n=-1} |  i = 0,1,2; l_{B,i} = 1,...,L_{B,i}\}$,
whose center position is  roughly located at the origin of  coordinate system.}

\textcolor{black}{Next, we need to analyze the motion laws of the bird to formulate its motion equations.} 
One of the main characteristics of bird motion is the   wing flapping, which can be formulated as a restricted rotation of scattering points in each wing around a corresponding rotation axis.
Let us take the $0$-th OFDM symbol time as the initial moment. 
 We  iteratively update  the wing flapping from the  CPC of bird to obtain the \emph{wing flapping update point cloud} (WFUPC) as $\{\widetilde{\mathbf{p}}_{l_{B,i},n} |  i = 0,1,2; l_{B,i} = 1,...,L_{B,i}\}$,  $n=0,...,N-1$. 
Specifically, 
assume that the two wings of a bird remain symmetrical.  
Denote
 the initial phase  and flapping  frequency  of the  wings as  $\beta_{B,i,0}$ and $F_{B}$,  respectively, 
and denote the wings' phase at the $n$-th symbol  as $\beta_{B,i,n}$.
Denote
the maximum and minimum angles that can be achieved during  wing flapping     as $\beta_{Bmax}$ and $\beta_{Bmin}$.
We manually designate two points for the $i$-th wing
 as the rotation axis of wing flapping, which are denoted as $\mathbf{p}^a_{B,i}$ and $\mathbf{p}^b_{B,i}$,  $i=1,2$.
In the same coordinate system, the rotation directions of the two wings are opposite.

Without loss of generality, we  only elaborate on the flapping process of  $\mathbb{S}_{B,1}$ here, whose flapping  frequency and   rotation axis direction are $F_B$ and $\mathbf{e}_{B,1} = \frac{\mathbf{p}^a_{B,1} - \mathbf{p}^b_{B,1}}{\left|\mathbf{p}^a_{B,1} - \mathbf{p}^b_{B,1}\right|_2}$.
When $n\!=\!0$, we  calculate the phase change of the wing as $\Delta \beta_{B,1,n} \!=\! \beta_{B,1,0}$.  When $n \! \neq \! 0$, we  calculate  $\Delta \beta_{B,1,n}$ as shown in Eq.~(18)  at the top of this page, where  $\beta_c \!=\! \beta_{Bmax}\!-\!\beta_{Bmin}\!>\!0$ and $\beta_{B,1,n}\!=\!\beta_{B,1,n-1} \!+\! \Delta \beta_{B,1,n-1}$. 
\textcolor{black}{Then we should rotate the point $\widetilde{\mathbf{p}}_{l_{B,1},n-1}$ around the   $1$-st rotation axis by angle $\Delta \beta_{B,1,n}$ to obtain the updated point $\widetilde{\mathbf{p}}_{l_{B,1},n}$.
According to  Rodrigues' rotation formula\cite{dai2015euler}, 
$\widetilde{\mathbf{p}}_{l_{B,1},n}$ can be 
 represented as}
\begin{equation}
	\begin{split}
		\begin{aligned}
			\label{deqn_ex1a}
\widetilde{\mathbf{p}}_{l_{B,1},n} & = \mathbf{p}^a_{B,1} + \mathbf{u}_{l_{B,1},n}\cos \Delta \beta_{B,1,n} \\& \quad+ (\mathbf{e}_{B,1} \times \mathbf{u}_{l_{B,1},n})\sin \Delta \beta_{B,1,n} \\& \quad + \mathbf{e}_{B,1} (\mathbf{e}_{B,1}^T \mathbf{u}_{l_{B,1},n})(1-\cos \Delta \beta_{B,1,n}),
		\end{aligned}
	\end{split}\tag{19}
\end{equation}
where $\mathbf{u}_{l_{B,1},n} = \widetilde{\mathbf{p}}_{l_{B,1},n-1} - \mathbf{p}^a_{B,1}$. Similarly, we can formulate $\widetilde{\mathbf{p}}_{l_{B,2},n}$. 
Meanwhile, since the body of the bird will not change in WFUPC, there is $\widetilde{\mathbf{p}}_{l_{B,0},n} = \widetilde{\mathbf{p}}_{l_{B,0},n-1}$.

Similar to the motion equations of the UAV, we also need to  formulate the bird's attitude update, position update, and coordinate transformation based on the bird's WFUPC.
Denote the initial  attitude and attitude jitter matrix of the bird as $[\gamma_{B,Y,0},\gamma_{B,P,0},\gamma_{B,R,0}]^T$ and $\mathbf{J}_{B} \in \mathbb{R}^{3\times 2}$, respectively,
and denote the  attitude at the $n$-th symbol  as 
$[\gamma_{B,Y,n},\gamma_{B,P,n},\gamma_{B,R,n}]^T$. 
Meanwhile, denote the initial position and motion velocity of the bird as
$\mathbf{p}_{B,0}$ and $\mathbf{v}_{B}$, respectively,  and  denote the position of the bird at the $n$-th symbol  as $\mathbf{p}_{B,n}$.
Note that the bird's attitude update, position update, and coordinate transformation are completely consistent with the UAV.
Therefore, 
Similar to Eq. (12) and Eq.~(16), we can ultimately obtain the multiple scattering points model of the bird as 
$\{ \mathbf{p}^{\star}_{l_{B,i},n} = ({r}_{l_{B,i},n}, {\theta}_{l_{B,i},n}, {\phi}_{l_{B,i},n}) |  i = 0,1,2; l_{B,i} = 1,...,L_{B,i}\}$, $n=0,...,N-1$.

Then based on  Eq.~(10), we can characterize the sensing echo channel matrix of  the bird on the  $m$-th subcarrier of the $n$-th OFDM symbol   as 
\begin{equation}
	\begin{split}
		\begin{aligned}
			\label{deqn_ex1a}
\mathbf{H}^{\rm bird}_{n,m} = \sum_{i=0}^{2}
 &  \sum_{l_{B,i}=1}^{L_{B,i}}
 \alpha_{l_{B,i}}  e^{\!-\!j\! \frac{4\pi\! f_m  {r}_{l_{B,i},n}}{c}}\!
\times  \\
& \mathbf{a}_{\!R}(\Psi_{l_{B,i},n},\Omega_{l_{B,i},n})
			\mathbf{a}^T_{\!H}(\Psi_{l_{B,i},n},\Omega_{l_{B,i},n}),
		\end{aligned}
	\end{split}\tag{20}
\end{equation}
where $\Psi_{l_{B,i},n} \!=\! \cos {\phi}_{l_{B,i},n} \cos {\theta}_{l_{B,i},n}$ and $\Omega_{l_{B,i},n} \!=\! \sin {\phi}_{l_{B,i},n}$.

\subsection{Echo Signals of  UAVs and Birds}

Assuming that the BS performs beamforming through HU-UPA. The transmission signal on the $m$-th subcarrier of the $n$-th OFDM symbol   is
\begin{equation}
	\begin{split}
		\begin{aligned}
			\label{deqn_ex1a}
{\mathbf{x}}_{n,m} =  \sqrt{\frac{{\rho}_{s}P_t}{N_H}}\mathbf{a}_{H}\left(\Psi_t,\Omega_t\right)s_{n,m},
		\end{aligned}
	\end{split}\tag{21}
\end{equation}
where $P_t$ is the transmission power of BS, ${\rho}_{s}$  is the power allocation factor towards the sensing direction, 
 $\left(\Psi_t,\Omega_t\right)$ is the beamforming direction of HU-UPA, and $s_{n,m}$ is the sensing detection signal. Meanwhile, assume that   RU-UPA of the BS also performs receive beamforming with $\mathbf{w}_{n,m} = \sqrt{\frac{1}{N_R}} \mathbf{a}_{R}\left(\Psi_r,\Omega_r\right)$. 
Then the frequency-domain echo signal received by  BS on the $m$-th subcarrier of the $n$-th symbol can be represented as
\begin{equation}
	\begin{split}
		\begin{aligned}
			\label{deqn_ex1a}
y_{n,m} = \mathbf{w}_{n,m}^H \mathbf{H}_{n,m}^{\rm target} {\mathbf{x}}_{n,m}^* + n_{n,m},
		\end{aligned}
	\end{split}\tag{22}
\end{equation}
where $\mathbf{H}_{n,m}^{\rm target} = \mathbf{H}_{n,m}^{\rm UAV}$ when the target is a UAV,  while  $\mathbf{H}_{n,m}^{\rm target} = \mathbf{H}_{n,m}^{\rm bird}$ when the target is a bird.
Moreover, $n_{n,m}$ is  zero-mean additive   Gaussian   noise  with variance  $\sigma^2$.

We command the transmitting  beam direction and receiving beam direction of the BS to directly point towards the initial direction of the UAV or the  bird. After receiving the echo signal $y_{n,m}$, we  erase the influence of $s_{n,m}$ and obtain
\begin{equation}
	\begin{split}
		\begin{aligned}
			\label{deqn_ex1a}
\check{y}_{n,m} = y_{n,m}/s_{n,m}.
		\end{aligned}
	\end{split}\tag{23}
\end{equation}
Then we concatenate the  echo signals on  all subcarriers and  all OFDM symbols into a matrix $\mathbf{Y} \in \mathbb{C}^{N\times M}$, where $\mathbf{Y}[n,m] = y_{n,m}$, $n=0,...,N-1$, and $m=0,...,M-1$.

\section{Dual Feature Fusion Enabled Low-Altitude Target Recognition Scheme}

In this section, we  develop the low-altitude target feature extraction method  to obtain the cmD spectrum and HRRP of the target, and then design a dual feature fusion enabled  low-altitude target recognition network with CNN to distinguish between UAVs and birds.

\begin{figure*}[!t]
	\centering
	\subfloat[]{\includegraphics[width=60mm]{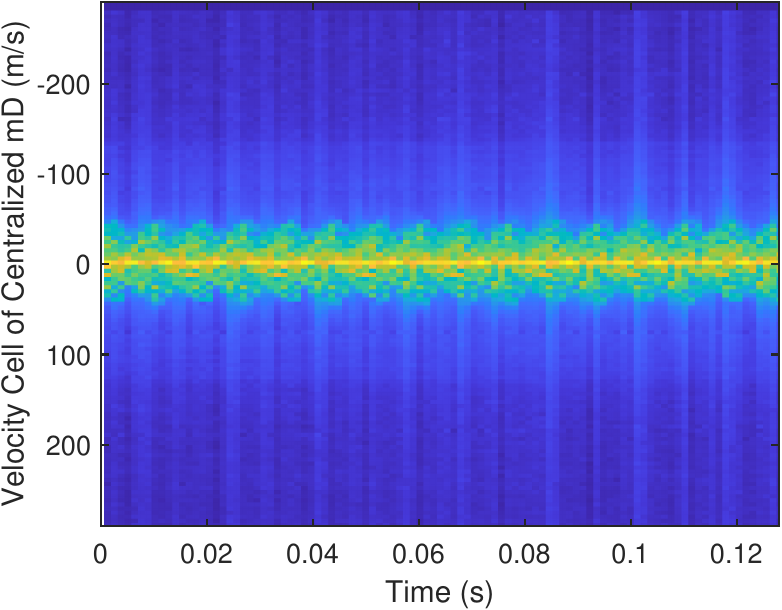}%
		\label{fig_first_case}}
	\hfil
	\subfloat[]{\includegraphics[width=60mm]{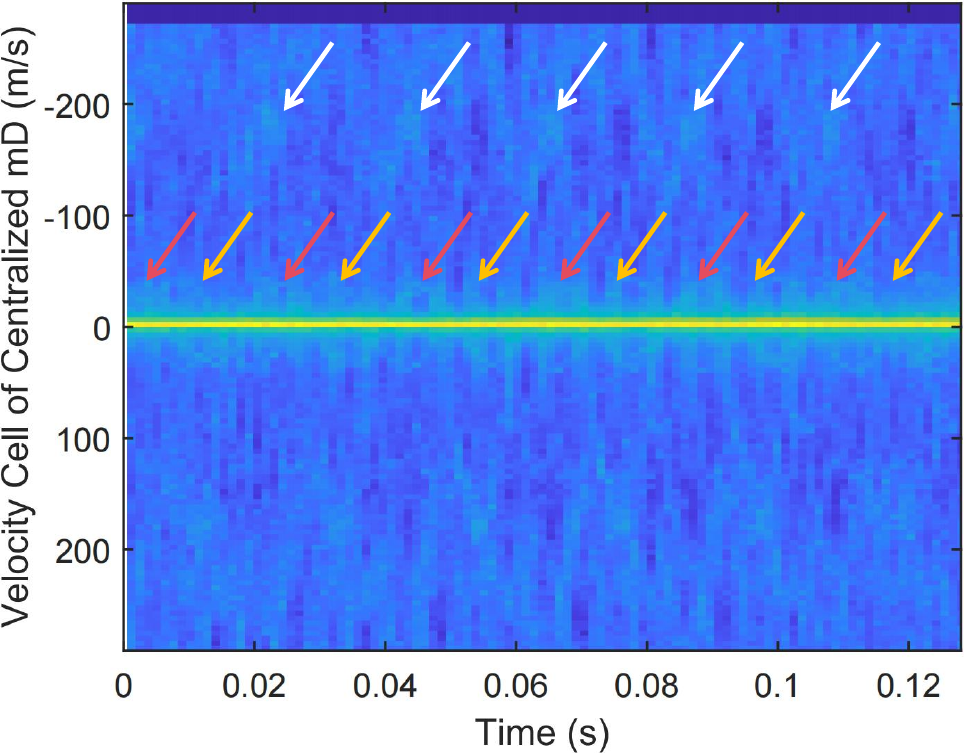}%
		\label{fig_first_case}}
	\hfil
	\subfloat[]{\includegraphics[width=60mm]{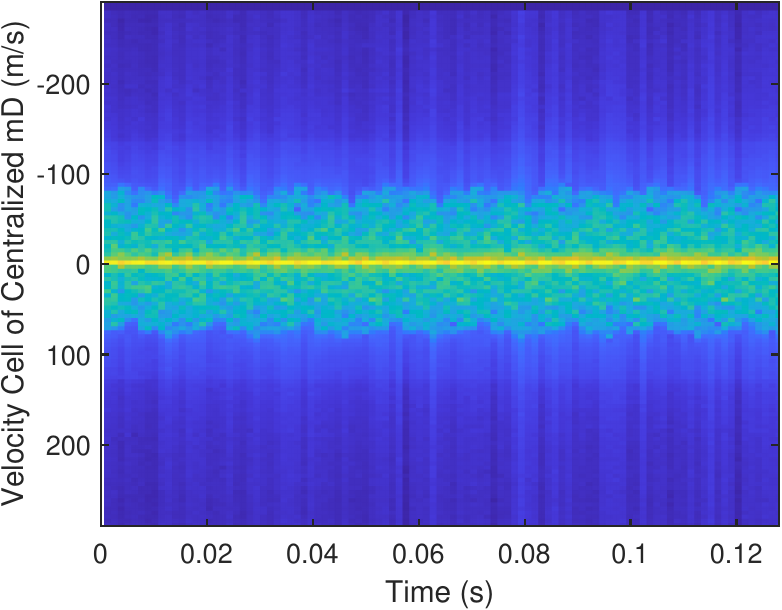}%
		\label{fig_first_case}}
	\hfil
	\subfloat[]{\includegraphics[width=60mm]{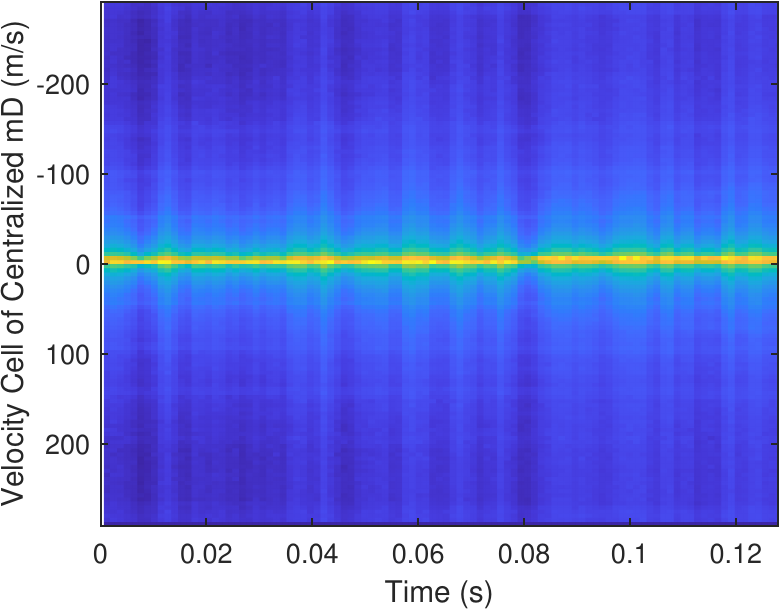}%
		\label{fig_first_case}}
	\hfil
	\subfloat[]{\includegraphics[width=60mm]{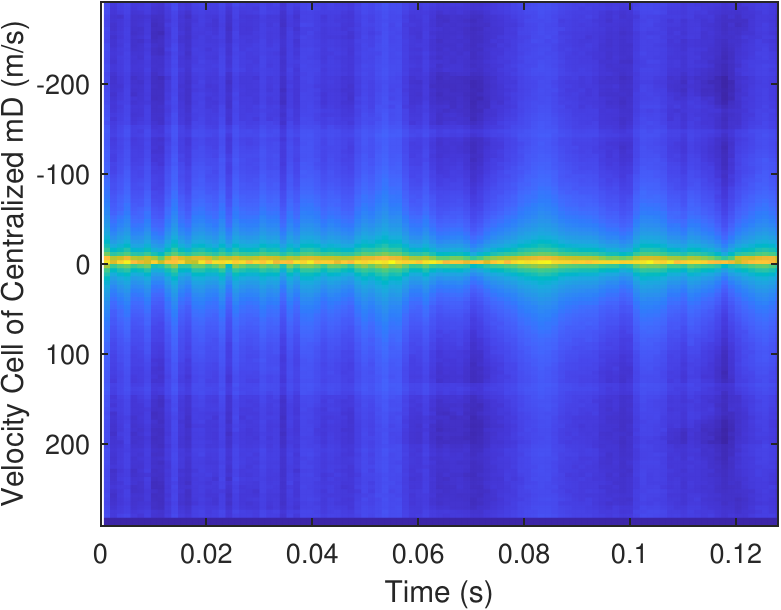}%
		\label{fig_first_case}}
	\hfil
	\subfloat[]{\includegraphics[width=60mm]{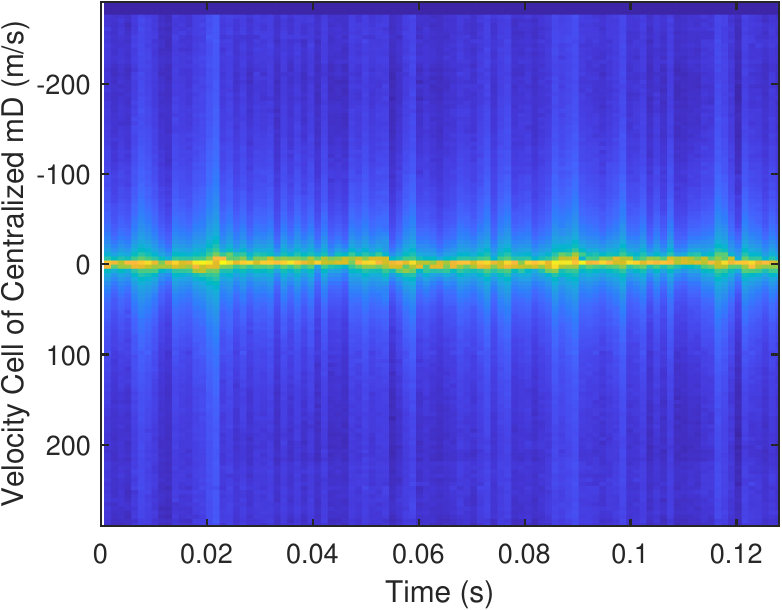}%
		\label{fig_second_case}}
	\caption{\textcolor{black}{
Centralized micro-Doppler spectrum of six different targets.
(a) DJI MAVIC 3, a four rotor UAV.
(b) Ehang eVTOL, an eight rotor UAV.
(c) Six Rotor UAV, a six rotor UAV.
(d) Crow.
(e) Pigeon.
(f) Sparrow. $f_0 = 26$ GHz, $\Delta f=480$ kHz}, $M = 4096$, $T_s = 10$ us, $N = 12800$, $N_0 = 128$, $G = 100$.}
	\label{fig_sim}
\end{figure*}

\subsection{cmD Spectrum of Low-Altitude Target}

Let us divide $N$ OFDM symbols into $G$ groups, each  containing $N_0$ symbols with $G$, $N_0$, $N \in \mathbb{N}^+$ and $N=GN_0$. 
We then   extract the submatrix corresponding to the $g$-th group from the echo signals $\mathbf{Y}$ as $\mathbf{Y}_g = \mathbf{Y}[(g-1)N_0+1:gN_0,:] \in \mathbb{C}^{N_0 \times M}$ with $g = 1,2,...,G$.
It can be inferred from Eq.~(17), Eq.~(20), Eq.~(22) and Eq.~(23) that $\mathbf{Y}_g$ contains the distance and velocity information for all scattering points of the low-altitude target.
Hence we perform  velocity DFT on $\mathbf{Y}_g$, i.e., perform $N_0$ points DFT  on each column of $\mathbf{Y}_g$ to obtain $\mathbf{Y}_{g,a}$ as
\begin{equation}
	\begin{split}
		\begin{aligned}
			\label{deqn_ex1a}
\mathbf{Y}_{g,a}[n_0,m] = \sum_{i_0 = 0}^{N_0-1}\mathbf{Y}_g[i_0,m]e^{-j\frac{2\pi i_0}{N_0}n_0},
		\end{aligned}
	\end{split}\tag{24}
\end{equation}
with $n_0 = 0,...,N_0-1$ and $m=0,...,M-1$.
Since the target's velocity possesses both positive and negative values, we move the zero frequency component of each column in $\mathbf{Y}_{g,a}$ to the center of each column, and obtain 
\begin{equation}
	\begin{split}
		\begin{aligned}
			\label{deqn_ex1a}
\mathbf{Y}_{g,b}[n_0,m] = \mathbf{Y}_{g,a}[(n_0+\lfloor N_0/2 \rfloor)  \ {\rm mod} \ N_0,m],
		\end{aligned}
	\end{split}\tag{25}
\end{equation}
where $\lfloor \cdot \rfloor$ represents the floor function, and 
$ {\rm mod} $ represents the  remainder function.
Then we perform distance inverse discrete Fourier transform (IDFT) on $\mathbf{Y}_{g,b}$, that is,   perform $M$ point IDFT on each row of $\mathbf{Y}_{g,b}$ to obtain 
the distance-velocity spectrum as 
\begin{equation}
	\begin{split}
		\begin{aligned}
			\label{deqn_ex1a}
\mathbf{Y}_{g,c}[n_0,m_0] =  \frac{1}{M}\sum_{i_0 = 0}^{M-1}\mathbf{Y}_{g,b}[n_0,i_0]e^{j\frac{2\pi i_0}{M}m_0},
		\end{aligned}
	\end{split}\tag{26}
\end{equation}
with $m_0 = 0,...M-1$ and  $n_0 = 0,...,N_0-1$.
Next, we take the modulus values of $\mathbf{Y}_{g,c}$ to obtain $\mathbf{Y}_{g,d}$ with 
$\mathbf{Y}_{g,d}[n_0,m_0] = 
{\rm abs}\{\mathbf{Y}_{g,c}[n_0,m_0]\}$, where 
${\rm abs}\{ \cdot \}$ represents taking the {modulus value}.
We   sum up all the elements in each row  of 
$\mathbf{Y}_{g,d}$ to obtain the mD spectrum slice for the $g$-th time period as 
\begin{equation}
	\begin{split}
		\begin{aligned}
			\label{deqn_ex1a}
\mathbf{y}_{g, \rm mD} = \sum_{m = 0}^{M-1}  \mathbf{Y}_{g,d}[:,m]  \in \mathbb{R}^{N_0 \times 1},
		\end{aligned}
	\end{split}\tag{27}
\end{equation}
which contains all the mD information of the low-altitude targets from the $(g-1)N_0$-th symbol  to the $(gN_0-1)$-th symbol.
Then we concatenate the mD spectrum slices of all $G$ groups together to obtain the mD spectrum matrix  as 
\begin{equation}\color{black}
	\begin{split}
		\begin{aligned}
			\label{deqn_ex1a}
\mathbf{Y}_{\rm mD} = [\mathbf{y}_{1, \rm mD}, \mathbf{y}_{2, \rm mD},...,\mathbf{y}_{G, \rm mD}] \in \mathbb{R}^{N_0 \times G}.
		\end{aligned}
	\end{split}\tag{28}
\end{equation}
Note that  different  velocities of the target will affect the central index of the mD spectrum, which in turn affect the generalization of subsequent target recognition. Hence we shift the Doppler unit where the target's velocity is located to the center row of $\mathbf{Y}_{\rm mD}$. For this purpose, let us calculate the index of the Doppler unit where the target velocity is located as $n_0^\star = \mathop{\arg\max}\limits_{n_0} \sum_{g=1}^G \mathbf{y}_{g,\rm mD}$. Then the {cmD spectrum} matrix can be represented as 
\begin{equation}
	\begin{split}
		\begin{aligned}
			\label{deqn_ex1a}
\mathbf{Y}_{\rm mD}^{\rm center} \! =  \! \frac{\mathbf{Y}_{\rm mD}^{\rm extend}\left [n_0^\star -\! \left \lfloor \! \frac{N_0}{2} \! \right \rfloor + 1:n_0^\star + \! \left \lfloor \! \frac{N_0}{2} \! \right \rfloor, : \right ]}{\max \{ \mathbf{Y}_{\rm mD} \}} \!  \in \mathbb{R}^{N_0 \times G},
		\end{aligned}
	\end{split}\tag{29}
\end{equation}
where $\mathbf{Y}_{\rm mD}^{\rm extend} = \left[\mathbf{0}_{N_0 \times G}^T,\mathbf{Y}_{\rm mD}^T, \mathbf{0}_{N_0 \times G}^T\right]^T \in \mathbf{R}^{3N_0 \times G}$, and ${\rm max}\{\cdot\}$ represents the function of finding the maximum value.

\begin{figure*}[!t]
	\centering
	\subfloat[]{\includegraphics[width=60mm]{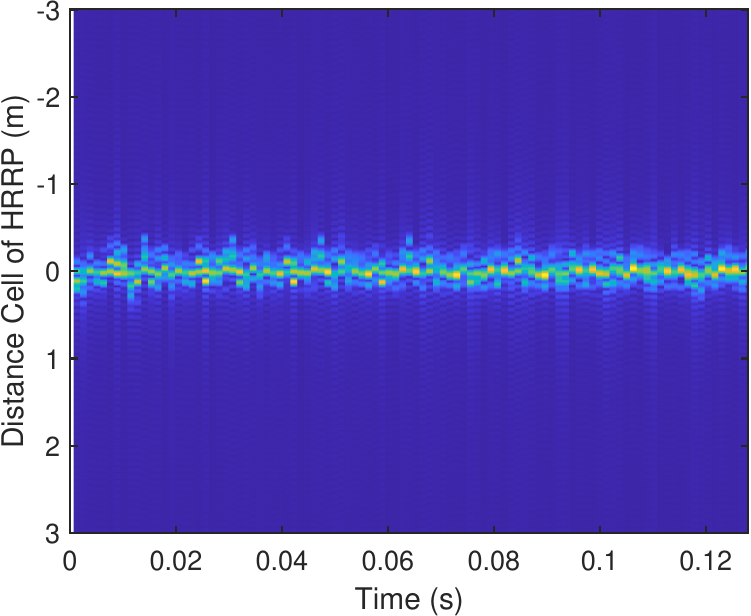}%
		\label{fig_first_case}}
	\hfil
	\subfloat[]{\includegraphics[width=60mm]{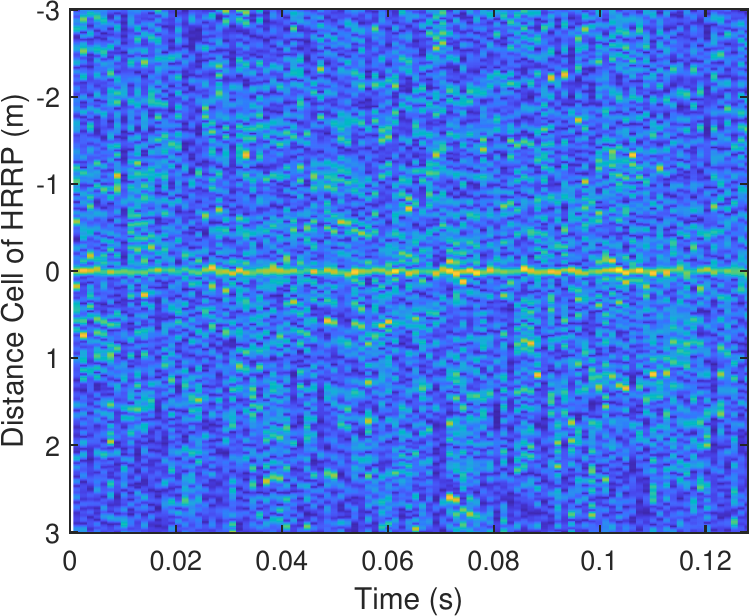}%
		\label{fig_first_case}}
	\hfil
	\subfloat[]{\includegraphics[width=60mm]{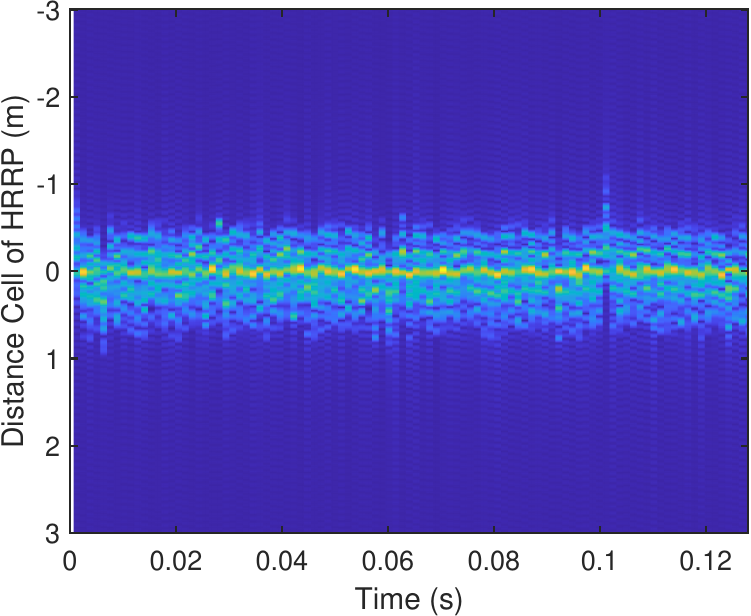}%
		\label{fig_first_case}}
	\hfil
	\subfloat[]{\includegraphics[width=60mm]{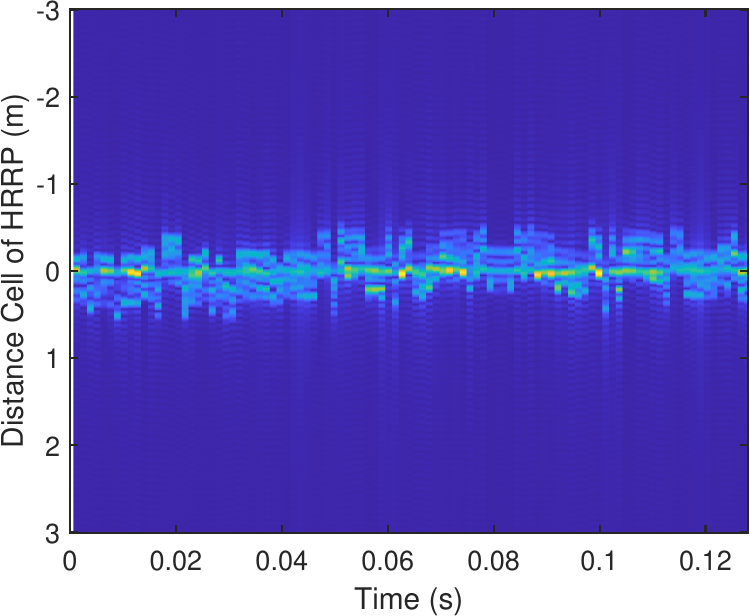}%
		\label{fig_first_case}}
	\hfil
	\subfloat[]{\includegraphics[width=60mm]{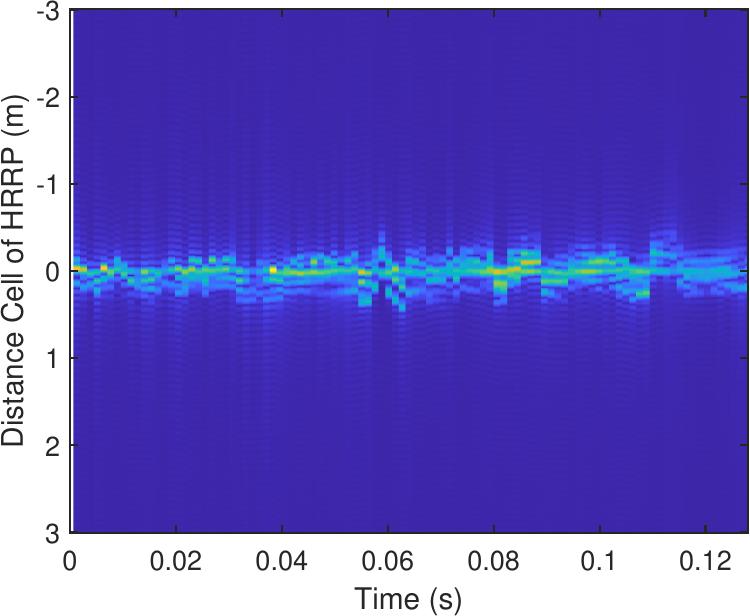}%
		\label{fig_first_case}}
	\hfil
	\subfloat[]{\includegraphics[width=60mm]{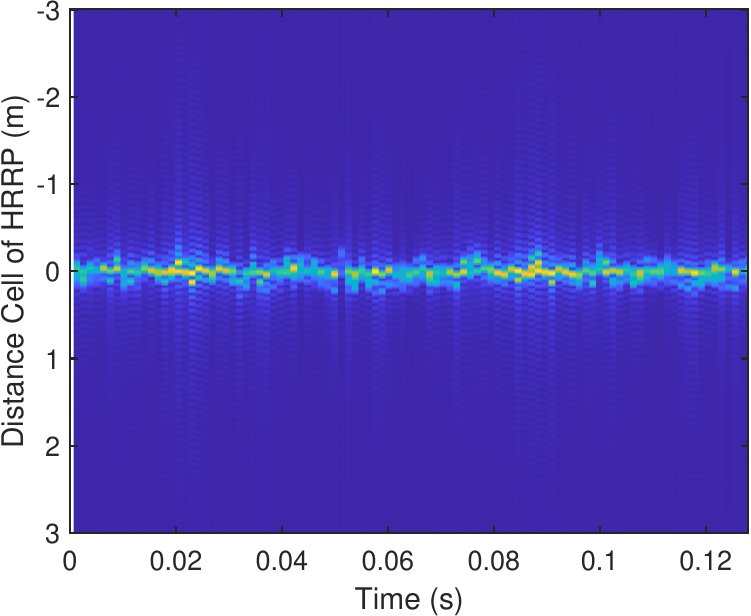}%
		\label{fig_second_case}}
	\caption{\textcolor{black}{
HRRP temporal sequence of six different  targets.
(a) DJI MAVIC 3.
(b) Ehang eVTOL.
(c) Six Rotor UAV. 
(d) Crow.
(e) Pigeon.
(f) Sparrow. $f_0 = 26$ GHz, $\Delta f=480$ kHz}, $M = 4096$, $T_s = 10$ us, $N = 12800$, $N_0 = 128$, $G = 100$, $[-r_{\rm HRRP,min},r_{\rm HRRP,max}] = [-3m,3m]$, $\Delta r_{\rm HRRP} = 0.01m$.}
	\label{fig_sim}
\end{figure*}

Fig.~5 shows the cmD spectrum of six different low-altitude targets, 
 including DJI MAVIC 3,
 Ehang eVTOL,
 Six Rotor UAV,
 crow,
 pigeon, and 
sparrow. 
{The specific information of these low-altitude targets can be seen in Fig.~8 in the subsequent simulations section.}
It can be seen from Fig.~5(a) and Fig.~5(c) that the cmD spectrums of the four rotor UAV   and the six rotor UAV  exhibit    obvious sine envelope, which corresponds to the high frequency rotation of the UAVs' paddles.
Besides, since the size of Ehang eVTOL is much larger than that of  typical UAVs, the multi-layered sine envelopes appear in the cmD spectrum as  shown in  Fig.~5(b), which is different from  other UAVs.
In addition,  it can be seen from Fig.~5(d), Fig.~5(e) and Fig.~5(f)  that only a sinusoidal envelope with slower frequency and lower amplitude can be seen in the cmD spectrum of the birds within a limited observation time, and even the sinusoidal envelope cannot be seen in the cmD spectrum of most birds.
This phenomenon is mainly due to  that the frequency of bird flapping wings is  relatively low.
Therefore, we can save the cmD spectrum of low-altitude targets as images for subsequent target  recognition.

\begin{figure*}[!t]
	\centering
	\includegraphics[width=180mm]{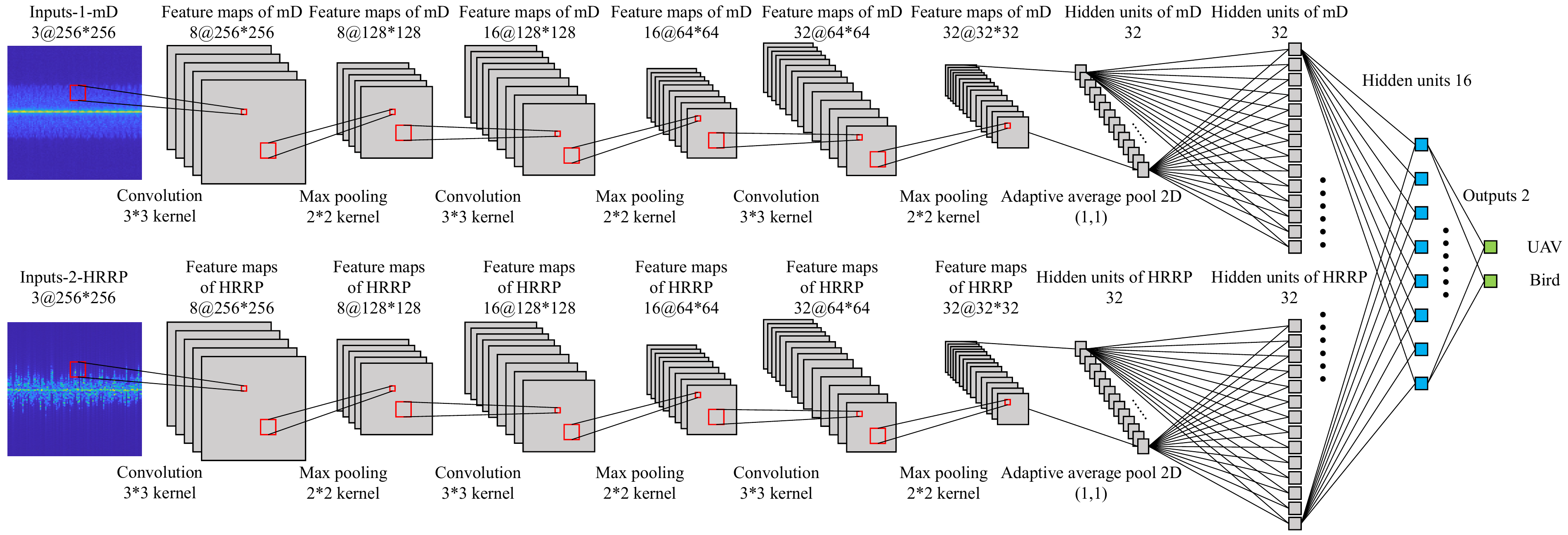}
	\caption{Low-altitude target recognition network based on convolutional neural network combined with micro-Doppler spectrum and HRRP.}
	\label{fig_1}
\end{figure*}

\subsection{HRRP Sequence Spectrum of Low-Altitude Target}

To obtain the HRRP of low-altitude target with low complexity, we extract the echo signal vector of the target at the $((g-1)N_0+1)$-th OFDM symbol from $\mathbf{Y}$ as $\mathbf{y}_{g} = \{\mathbf{Y}[(g-1)N_0+1,:]\}^T \in \mathbb{C}^{M\times 1}$, where $g=1,...,G$. 
Then we perform $M$ point distance IDFT on $\mathbf{y}_{g}$ to obtain 
 $\mathbf{y}_{g,a}$ with 
\begin{equation}
	\begin{split}
		\begin{aligned}
			\label{deqn_ex1a}
\mathbf{y}_{g,a}[m_0] =  \frac{1}{M}\sum_{i_0 = 0}^{M-1}\mathbf{y}_{g}[i_0]e^{j\frac{2\pi i_0}{M}m_0},
		\end{aligned}
	\end{split}\tag{30}
\end{equation}
where $m_0 = 0,...,M-1$.
Then we take the {modulus values} of $\mathbf{y}_{g,a}$ to obtain $\mathbf{y}_{g,b}$ with $\mathbf{Y}_{g,b}[m_0] = {\rm abs}\{ \mathbf{Y}_{g,a}[m_0]\}$. 
By calculating the index number of the maximum value in $\mathbf{y}_{g,b}$ as $m_{0,g}^\star = \mathop{\arg\max}\limits_{m_0} \mathbf{y}_{g,b}$, we can obtain the distance estimation result of the target in the $g$-th group as 
$\hat{r}_g = \frac{c(m_{0,g}^\star-1)}{2W}$. 
Then we construct a distance search vector as $\mathbf{r}_g = (\hat{r}_g-r_{\rm HRRP,min}:\Delta r_{\rm HRRP}:\hat{r}_g+r_{\rm HRRP,max})^T$, where $(\zeta_1:\Delta \zeta: \zeta_2)^T$ represents constructing a column vector with $\zeta_1$ as the initial value, $\Delta \zeta$ as the step size, and $\zeta_2$ as the termination value.
Moreover, 
$\Delta r_{\rm HRRP}$ and $[-r_{\rm HRRP,min},r_{\rm HRRP,max}]$ are the resolution and range of HRRP set by the system, respectively.
Denote the length of $\mathbf{r}_g$ as ${\rm len}(\mathbf{r}_g)$, and there should be ${\rm len}(\mathbf{r}_g) \equiv \frac{r_{\rm HRRP,max}+r_{\rm HRRP,min}+\Delta r_{\rm HRRP}}{\Delta r_{\rm HRRP}} \triangleq I$.
For each element $\mathbf{r}_g[i]$ in $\mathbf{r}_g$, we construct a dictionary vector as $\mathbf{a}_{g,i} \in \mathbb{C}^{M\times 1}$ with
$\mathbf{a}_{g,i}[m] = e^{-j4\pi \frac{f_m \mathbf{r}_g[i]}{c}}$, where $i=1,...,I$ and $m=0,...,M-1$. Let us calculate the matching value corresponding to $\mathbf{r}_g[i]$ as 
\begin{equation}
	\begin{split}
		\begin{aligned}
			\label{deqn_ex1a}
G_{g,i} =    {\rm abs} \{   \mathbf{a}_{g,i}^H   \mathbf{y}_{g,a} \}.
		\end{aligned}
	\end{split}\tag{31}
\end{equation}
Then  we stack   $\{ G_{g,i}| i = 1,...,I; g = 1,...,G \}$  into a matrix and obtain the  HRRP sequence spectrum matrix of the low-altitude target as 
$\mathbf{Y}_{\rm HRRP} \in \mathbb{R}^{I \times G}$ with $\mathbf{Y}_{\rm HRRP}[i,g] = G_{g,i}$. 

Fig.~6 shows the HRRP sequence spectrum of six different low-altitude targets, also  including DJI MAVIC 3,
 Ehang eVTOL,
 Six Rotor UAV,
 crow,
 pigeon, and 
sparrow. 
 It is seen  from Fig.~6(b) that Ehang eVTOL shows significant broadening throughout the entire HRRP range of $[-3m, 3m]$.
The reason is  that the size of Ehang eVTOL is much larger than other UAVs and birds.  
Besides, since the rotation frequency of the UAV's paddles is much higher than the wing frequency of ordinary birds, the HRRP broadening of the six rotor UAV is more continuous as can be seen from  Fig.~6(c), Fig.~6(d), and Fig.~6(e), which will help  distinguish between UAVs and birds.
From Fig.~6(a), Fig.~6(d), and Fig.~6(e), it can be found that the HRRP of DJI MAVIC 3 possesses more obvious periodicity than that of  crow and pigeon. However, this phenomenon is not obvious for DJI MAVIC 3 and sparrow. Therefore, although HRRP can help distinguish between UAVs and birds, the gain   may be  limited for some specific low-altitude targets.

\subsection{Low-Altitude Target Recognition  Network}

We next design a dual feature fusion enabled low-altitude target recognition network with CNN as shown in Fig.~7.
The inputs of the low-altitude target recognition network are two RGB color images with dimensions of $3$@$256 \times 256$.
Here ``$3$@'' represents the number of channels of the image, and ``$256 \times 256$''  represents the data dimensions in each image channel.  
These two images respectively depict the visualization results of the target's cmD spectrum and HRRP sequence spectrum.
The cmD image and HRRP image are processed through three convolutional layers with the kernel size  $3 \times 3$ and three max-pooling layers with the kernel size  $2 \times 2$, respectively, to obtain the deep feature maps  with the size  $32$@$32\times 32$. Then we utilize the adaptive average pooling to transform the deep feature maps into $32$ hidden units, which  are  then transformed into $32$ new hidden units through a fully connected layer. Next we concatenate the hidden units of the cmD image and HRRP image into $64$ hidden units, and use two fully connected layers to ultimately transform these hidden units into an output layer with $2$ units.
In the calculation process, the activation functions of the convolutional layer and fully connected layer are Rectified Linear Unit (ReLU) functions, while the activation function of the output layer is Softmax function, which will be beneficial for the  recognition  of UAV and bird\cite{lecun1998gradient}.
The designed low-altitude target recognition network will use a large amount of cmD and HRRP  image samples for offline training in a data-driven manner. The training process will employ the cross entropy loss function and the adaptive moment estimation (Adam) optimizer. Once the network training converges, the best model will be deployed for online testing.

\subsection{\textcolor{black}{Computational Complexity Analysis}}

\textcolor{black}{
The computational complexity of the proposed scheme mainly comes from low-altitude target feature extraction and AI recognition network. 
In terms of feature extraction, 
the computational complexity of cmD feature extraction is 
$O\{ G ( MN_0^2 + M^2 N_0 ) \}$, and 
the computational complexity of HRRP feature extraction is
$O\{ G (M^2+MI) \}$.  
Different from traditional signal processing, the computational complexity of AI network is typically represented by the number of model parameters. 
Here the number of parameters for the proposed low-altitude target recognition network is $0.15$ M, which is a lightweight network and  easy to deploy on edge devices.}

\section{Simulation Results}

{
In this section, we  generate a UAV and bird echo signal dataset, and then evaluate the performance of the proposed low-altitude target recognition scheme.}

\subsection{{UAV and Bird Echo Signal Dataset}}

\begin{figure*}[!t]
	\centering
	\includegraphics[width=180mm]{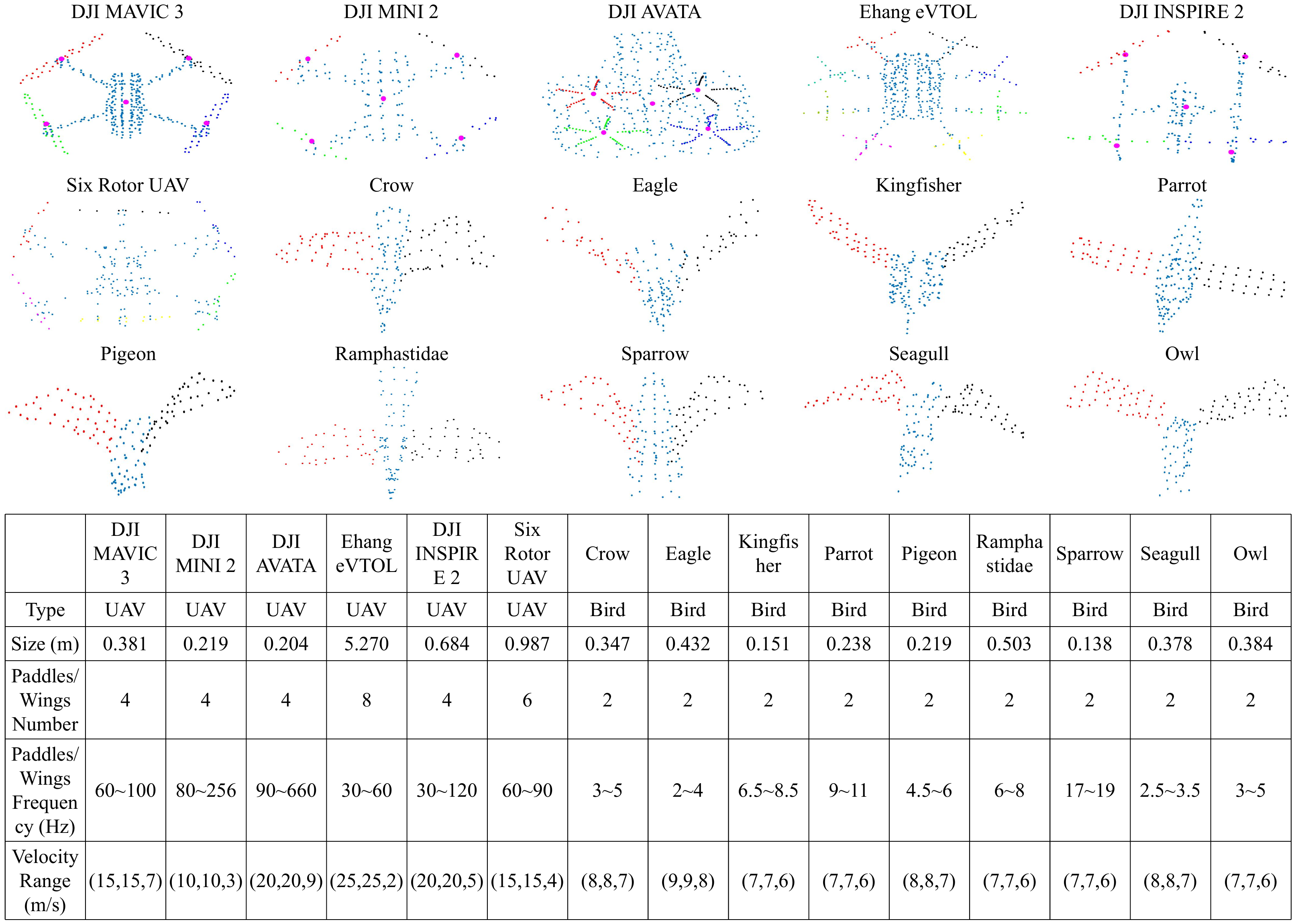}
	\caption{Calibration point clouds  and main information for $15$ types of low-altitude targets.}
	\label{fig_1}
\end{figure*}

{Based on the echo signal calculation process in Section~\RNum{3},} we  first construct the  CPCs for $6$ types of UAVs and $9$ types of birds.
The UAVs include DJI MAVIC 3, DJI MINI~2, DJI AVATA, Ehang eVTOL, DJI INSPIRE 2, and  Six Rotor UAV.
\textcolor{black}{Among them, DJI MAVIC 3, DJI MINI 2, DJI AVATA, and DJI INSPIRE 2 are four  rotor UAVs; Six Rotor UAV is a six rotor UAV; and Ehang eVTOL is a eight rotor UAV.}
The birds 
include crow, eagle, kingfisher, parrot, pigeon, ramphastidae, sparrow, seagull, and owl. The CPCs and main information of these low-altitude targets are shown in Fig.~8\footnote{\textcolor{black}{The  information of the birds is mainly obtained through \cite{7458568499999, asasas, ggggg5555}, and  the  information of the UAVs is mainly obtained through \cite{9764172, dji888}.}}, in which 
the size of a UAV refers to the length of the diagonal of its body,
the size of a bird refers to the length of its body from head to toe,
the paddle frequency refers to the number of rotations of the UAV's paddle within $1$ s,  the  wing frequency refers to the number of times a bird flaps its wings within $1$ s,
and the velocity range refers to the absolute value of the 3D velocity  not exceeding $(v_x,v_y,v_z)$.   

{To generate the UAV and bird echo signal dataset,}  we set the operating frequency of  BS as $f_0 = 26$~GHz, set the subcarrier spacing as $\Delta f = 480$ kHz,
set the number of subcarriers as $M = 4096$, set the antenna spacing as $d_x = d_z = d = \frac{\lambda}{2} = \frac{c}{2f_0}$, set the array size of HU-UPA as $N^{x}_{H}\times N_{H}^z = 8 \times 8$, set the array size of RU-UPA as $N^{x}_{R}\times N_{R}^z = 8 \times 8$, and set the OFDM symbol interval as $T_s = 10$ us. 
To fully extract the cmD and HRRP features of low-altitude target, we set the number of OFDM symbols contained in one \textcolor{black}{TCF} as $N = 12800$, and then the corresponding total observation duration is $0.128$ s.

{
With these system parameters and CPCs,} 
 we generate $300$ motion models corresponding to each type of UAV and $200$ motion models corresponding to each type of  bird by randomly changing the position, velocity, attitude, paddle frequency, wing frequency and other  parameters of UAVs and birds.
Then with these $3600$ motion models,
based on Eq. (17), Eq. (20), and Eq. (22), 
 we finally generate a UAV and bird echo signal dataset
containing 
 $3600$ noiseless echo signal  matrices.

\subsection{{Data Preprocessing and  Evaluation Metric for the Low-Altitude Target Recognition Network}}

{
For the data preprocessing of the low-altitude target recognition network,}
we add random noise with  SNR belongs to \{-20, -15, -10, -5, 0, 5, 10, 15, 20, 25, 30\}~dB to the noiseless echo signal  matrices, where each noiseless echo signal matrix independently repeats $6$  times at each SNR, and then $3600 \times 11 \times 6 = 237600$ noisy echo signal matrices are obtained. 
Here based on Eq.~(22), the SNR of echo signal can be defined as
\begin{equation}\color{black}
	\begin{split}
		\begin{aligned}
			\label{deqn_ex1a}
{\rm SNR} = \frac{\mathbb{E}\left\{|\mathbf{w}_{n,m}^H \mathbf{H}_{n,m}^{\rm target} {\mathbf{x}}_{n,m}^*|^2\right\}}{\mathbb{E}\left\{|n_{n,m}|^2\right\}}.
		\end{aligned}
	\end{split}\tag{32}
\end{equation}

Next,   we set the number of groups for feature extraction as
$G = 100$ with each group containing $N_0 = 128$ symbols.   We
set the observation range and resolution of HRRP as 
$[-r_{\rm HRRP,min},r_{\rm HRRP,max}] = [-3m,3m]$ and $\Delta r_{\rm HRRP} = 0.01m$.  
Then we extract one cmD  image and one HRRP  image from each noisy echo signal matrix, where the image size is set as $3$@$256\times 256$. To facilitate the storage and retrieval of the images, we concatenate the cmD image and HRRP image into an aggregated image with a size  $3$@$512\times 256$, which will be re-split into one mD image and one HRRP image during the data loading stage of the  network. Finally, we save these  $237600$ aggregated images  to train, validate, and evaluate the  low-altitude target recognition network.

Assuming there are a total of $Z_{U,0}$ UAV image samples and $Z_{B,0}$ bird image samples, the low-altitude target recognition network recognizes $Z_{U,1}$ out of all the UAV image samples as UAV, and recognizes $Z_{B,1}$ out of all the bird image samples as bird. Then the recognition accuracy of the network  can be defined as 
\begin{equation}\color{black}
	\begin{split}
		\begin{aligned}
			\label{deqn_ex1a}
{\rm Acc} = \frac{Z_{U,1} + Z_{B,1}}{Z_{U,0} + Z_{B,0}},
		\end{aligned}
	\end{split}\tag{33}
\end{equation}
which will be employed as the performance evaluation metric.

\begin{figure}[!t]
	\centering
	\includegraphics[width=80mm]{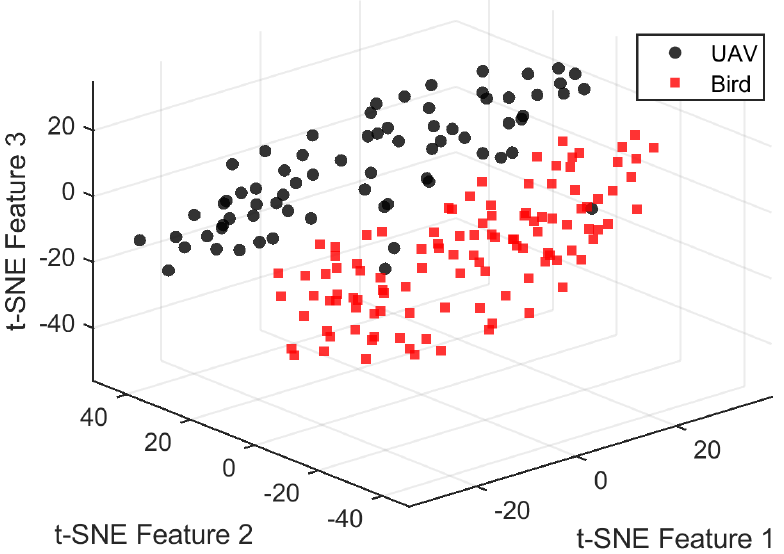}
	\caption{t-SNE analysis on $200$ randomly selected samples from   $10$ dB dataset.}
	\label{fig_1}
\end{figure}

\subsection{Analysis of t-Distributed Stochastic Neighbor Embedding for cmD and HRRP}

To verify the differences in cmD and HRRP features between UAVs and birds, we utilize the pre-trained ResNet-50 model to extract the deep level features from cmD and HRRP\cite{he2016deep}.
These deep level features extracted through convolutional layers and fully connected layers are high-dimensional vectors. 
We  then utilize the 
t-distributed stochastic neighbor embedding (t-SNE) to reduce these high-dimensional feature vectors to 3D vectors and show them within the 3D space in   Fig.~9\cite{NIPS2002_6150ccc6,9754679,van2008visualizing}.
It can be seen that the t-SNE feature map of these deep level features  exhibit good aggregation results, and hence 
cmD and HRRP can  be used for the recognition of UAVs and birds.

\begin{figure}[!t]
	\centering
	\subfloat[]{\includegraphics[width=80mm]{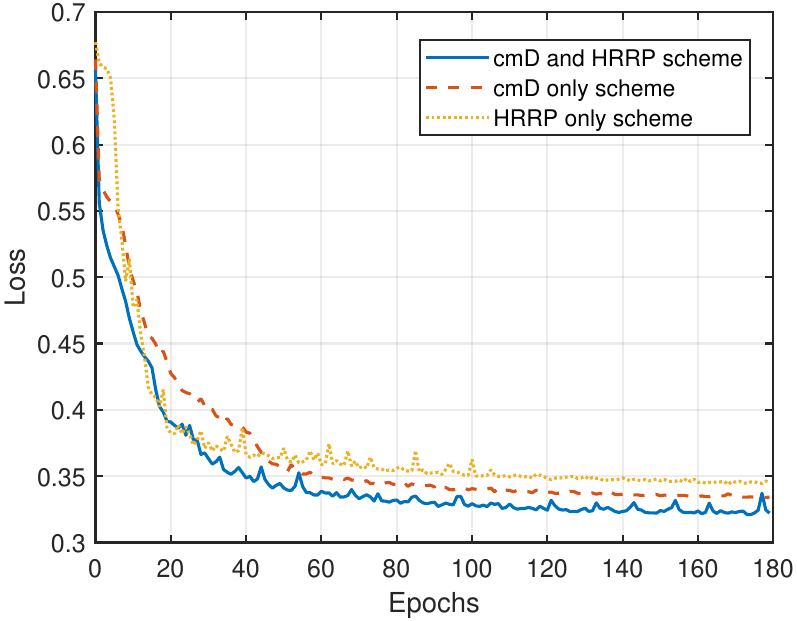}%
		\label{fig_first_case}}
	\vfil
	\subfloat[]{\includegraphics[width=79mm]{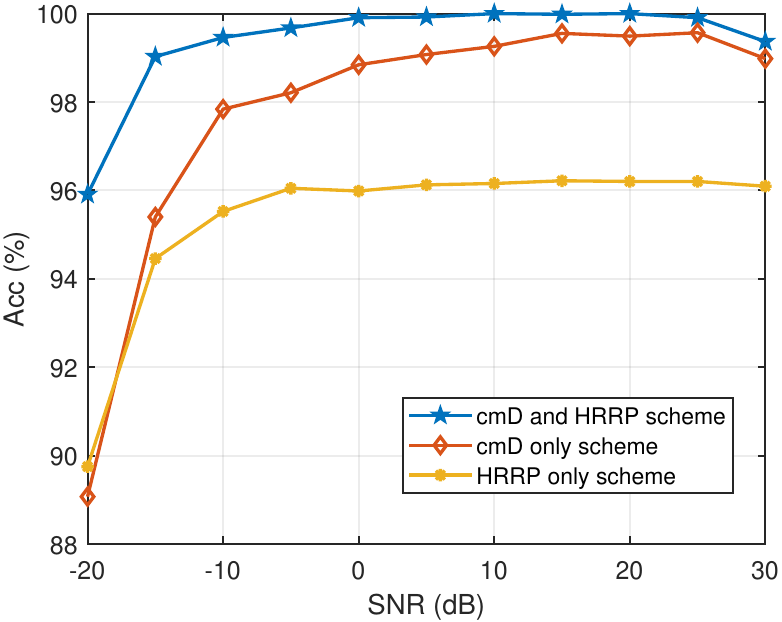}%
		\label{fig_first_case}}
	\caption{
Training loss and testing accuracy of low-altitude target recognition.
(a) The network training loss  versus training epochs of different schemes.
(b) The network testing recognition  accuracy versus SNR of different schemes.}
	\label{fig_sim}
\end{figure}

\subsection{Overall Accuracy of Low-Altitude Target Recognition}

To conduct feature comparison, we  refer to the scheme that combines cmD and HRRP features as ``cmD and HRRP scheme'', refer to the scheme that only uses cmD feature as ``cmD only scheme'', and refer to the scheme that only uses HRRP feature as ``HRRP only scheme''. Then we select 190080, 23760, and 23760 samples from all the aggregated images  as the training set, validation set, and testing set, respectively. 
Fig.~10 shows the training loss curve of the low-altitude target recognition network on the training set, as well as 
the  recognition  accuracy versus SNR on the testing set.

As shown in Fig.~10(a), the loss of network model  gradually decreases with the increase of training epochs and eventually tends to converge.
This means that the proposed low-altitude target recognition network has good convergence on dataset of UAVs and birds.
Besides, it can be seen that the convergence value of cmD and HRRP scheme is lower than those of cmD only scheme and HRRP only scheme,
which means that the cmD and HRRP scheme  has learned more comprehensive data features from the dataset.

As shown in Fig.~10(b), the overall recognition accuracy of the trained model on the testing set gradually improves with the increase of SNR. The average accuracy of cmD and HRRP scheme, cmD only scheme, and HRRP only scheme on all SNRs are 99.37\%, 97.75\%, and 95.34\%, respectively. 
The average accuracy of cmD and HRRP scheme, cmD only scheme, and HRRP only scheme for the samples below 0 dB are 98.51\%, 95.13\%, and 93.94\%, respectively.
These results confirm the effectiveness of the proposed low-altitude target recognition scheme.
Besides, the overall highest accuracy of the HRRP scheme can only reach 96.21\%, which means that the HRRP cannot  recognize some specific  targets. 
Meanwhile, it can be seen that in terms of overall recognition ability, the accuracy of the cmD and HRRP scheme is higher than that of the cmD only scheme, and the accuracy of the cmD only scheme is higher than that of the HRRP only scheme.

\begin{figure*}[!t]
	\centering
	\includegraphics[width=180mm]{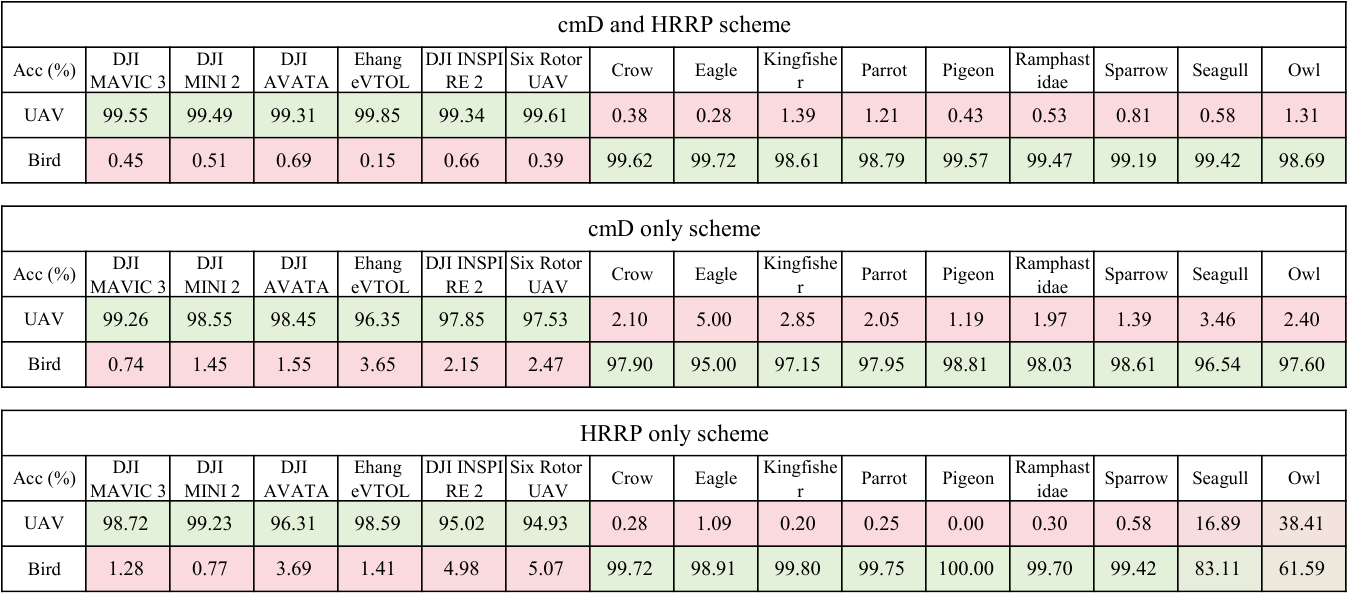}
	\caption{Confusion matrices of recognition accuracy for all types of UAVs and birds, in which the models used are the trained models in Section \RNum{5}-C.}
	\label{fig_1}
\end{figure*}

\begin{figure*}[!t]
	\centering
	\includegraphics[width=180mm]{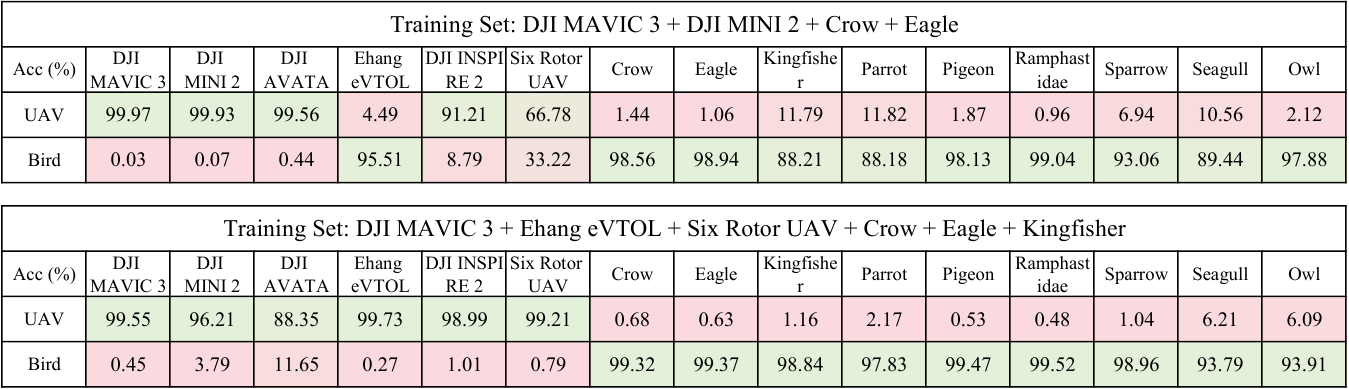}
	\caption{Confusion matrices of recognition accuracy obtained by training models with different datasets.}
	\label{fig_1}
\end{figure*}

\subsection{Accuracy for Different Types of UAVs and Birds}

Based on the models trained in Section \RNum{4}~C, we further test the recognition accuracy of the models on different types of UAVs and birds, and the confusion matrices are shown in Fig.~11.
It can be seen that the recognition accuracy of cmD and HRRP scheme on all types of low-altitude targets is above 98\%.
The cmD only scheme performs poorly in recognizing eagle to bird.
The HRRP only scheme performs poorly in recognizing DJI INSPIRE 2 to UAV, and performs poorly in recognizing seagull and owl to bird. 
Importantly, Fig.~11 indicates that combining cmD and HRRP can realize better recognition performance on most types of low-altitude targets than using them alone.

\vspace{-3mm}

\subsection{Impact of Datasets on Recognition Generalization}

We train the low-altitude target recognition network with different sub-datasets in  cmD and HRRP scheme, and then the confusion matrices obtained from the converged models on the testing set are shown in Fig.~12.
When the training set only includes two types of four rotor UAV,  
DJI MAVIC 3 and  DJI MINI 2, as well as two types of birds, crow and eagle, the recognition accuracy   for Six Rotor UAV and Ehang eVTOL  decrease to 66.78\% and 4.49\%. This indicates that the network model trained from the four rotor UAV cannot generalize to the six rotor UAV and Ehang eVTOL.
Hence Fig.~12 inspires that a dataset covering most types of UAVs and birds is necessary for training the low-altitude target recognition network.

\vspace{-3mm}

\section{Conclusions}

In this paper, we  proposed a UAV and bird recognition scheme for ISAC system.
{We    \textcolor{black}{formulated} the  motion equations and echo signals  for  UAVs and  birds.}
Then we  extracted the cmD spectrum and the HRRP of low-altitude targets from the echoes. 
Next, we  designed a  dual feature fusion enabled  low-altitude target recognition network with CNN, which employed the images of cmD  spectrum  and HRRP  as inputs to jointly recognize the categories of  targets. 
\textcolor{black}{Meanwhile,  we generated   237600 cmD and HRRP image samples  to train, validate, and evaluate the designed  low-altitude target recognition network.
The simulation results confirmed the effectiveness of the proposed scheme.}

\bibliographystyle{ieeetr}
\bibliography{ref.bib}

\vfill

\end{document}